\documentclass[preprint, onecolumn,preprintnumbers,tightenlines,epsfig,floatfix]{revtex4}
\usepackage{graphicx}
\usepackage{dcolumn}
\usepackage{bm}
\usepackage{amsfonts}
\usepackage{amssymb}
\usepackage{amsmath}

\newtheorem{theorem}{Theorem}
\newtheorem{acknowledgement}[theorem]{Acknowledgement}
\newtheorem{algorithm}[theorem]{Algorithm}

\newtheorem{definition}[theorem]{Definition}
\newtheorem{example}[theorem]{Example}

\newtheorem{notation}[theorem]{Notation}

\newtheorem{proof}[theorem]{Proof}

\begin{document}

\title{Quantum Discrete Cosine Transform for Image Compression}
\author{Chao-Yang Pang$^{1,2}$}
\email{cyp_900@hotmail.com}
\author{Zheng-Wei Zhou$^{1}$}
\email{zwzhou@ustc.edu.cn}
\author{Guang-Can Guo$^{1}$}
\email{gcguo@ustc.edu.cn}
\affiliation{Key Laboratory of Quantum Information, University of Science and Technology
of China, Chinese Academy of Sciences, Hefei, Anhui 230026, China$^{1}$\\
College of Mathematics and Software Science, Sichuan Normal University,
Chengdu, Sichuan 610066, People's Republic of China$^{2}$}

\begin{abstract}
Discrete Cosine Transform (DCT) is very important in image compression.
Classical 1-D DCT and 2-D DCT has time complexity $O(N\log _{2}N)$ and $%
O(N^{2}\log _{2}N)$ respectively. This paper presents a quantum DCT
iteration, and\ constructs a quantum 1-D and 2-D DCT algorithm for image
compression by using the iteration. The presented 1-D and 2-D DCT has time
complexity $O(\sqrt{N})$ and $O(N)$ respectively. In addition, the method
presented in this paper generalizes the famous Grover's algorithm to solve
complex unstructured search problem.
\end{abstract}

\maketitle


\section{Introduction}

\subsection{The Conception of Digital Image}

\markboth{ Submitted to IEEE Transactions On Information Theory on 6
January 2006}
{Murray and Balemi: Using the Document Class IEEEtran.cls}%
\setcounter{page}{1}The result of sampling and quantization of a monochromic
N-by-N image is a matrix of real numbers\cite{1,2}

\begin{equation}
F=\left[
\begin{array}{cccc}
f_{00} & f_{01} & ... & f_{0(N-1)} \\
f_{10} & f_{11} & ... & f_{1(N-1)} \\
\vdots & \vdots & \vdots & \vdots \\
f_{(N-1)0} & f_{(N-1)1} & ... & f_{(N-1)(N-1)}%
\end{array}%
\right]  \label{1}
\end{equation}

Each element of this matrix is called \emph{pixel}. The pixel $f_{ij}$ is a
gray-level value. The value $f_{ij}^{2}$ is proportional to brightness or
energy.

Let column vector

\begin{equation}
\overset{\rightarrow }{f_{i}}=(f_{0i},f_{1i},...,f_{(N-1)i})^{T}  \label{2}
\end{equation}

, where $T$ represents the transpose of a vector.

The digital image $F$ can be expressed as formula

\begin{equation}
F=\left[
\begin{array}{cccc}
\overset{\rightarrow }{f_{0}} & \overset{\rightarrow }{f_{1}} & ... &
\overrightarrow{f_{N-1}}%
\end{array}%
\right]  \label{3}
\end{equation}

Fig. 1 illustrates the conception of digital image.

\bigskip

\begin{tabular}{c}
{\Large Insert Fig1 Here}%
\end{tabular}

\bigskip

\subsection{Introduction of Classical Discrete Cosine Transform (DCT)}

\subsubsection{One-Dimensional Discrete Cosine Transform (1-D DCT)}

The one-dimensional discrete cosine transform\cite{1,2} (1-D DCT) and
inverse DCT are defined as

\begin{equation}
\begin{tabular}{c}
$\left\{
\begin{tabular}{c}
$c_{u}=\sum\limits_{n=0}^{N-1}\alpha _{u}f_{n}\cos \frac{(2n+1)u\pi }{2N}$
\\
$f_{n}=\sum\limits_{n=0}^{N-1}a_{u}c_{u}\cos \frac{(2n+1)u\pi }{2N}$%
\end{tabular}%
\right. $ \\
$(0\leq u,n\leq N-1)$%
\end{tabular}
\label{4}
\end{equation}

, where%
\begin{equation*}
\alpha _{u}=\left\{
\begin{tabular}{ccc}
$1/\sqrt{N}$ & $if$ & $u=0$ \\
$\sqrt{2/N}$ & $if$ & $1<u<N$%
\end{tabular}%
\right.
\end{equation*}

The $N\times N$ DCT matrix is

\begin{equation}
\begin{tabular}{l}
$D=(\alpha _{u}\cos \frac{(2n+1)\times u\pi }{2N})_{N\times N}=$ \\
$\left[
\begin{tabular}{ccc}
$\alpha _{0}$ & $\cdots $ & $\alpha _{0}$ \\
$\alpha _{1}\cos \frac{1\times 1\pi }{2N}$ & $\cdots $ & $\alpha _{1}\cos
\frac{(2N-1)\times 1\pi }{2N}$ \\
$\vdots $ & $\cdots $ & $\vdots $ \\
$\alpha _{N-1}\cos \frac{1\times (N-1)\pi }{2N}$ & $\cdots $ & $\alpha
_{N-1}\cos \frac{(2N-1)\times (N-1)\pi }{2N}$%
\end{tabular}%
\right] $%
\end{tabular}
\label{5}
\end{equation}

Let vector

\begin{equation}
\overset{\rightarrow }{D}_{u}=\alpha _{u}(\cos \frac{1.u\pi }{2N},\cos \frac{%
3.u\pi }{2N},...,\cos \frac{(2N-1).u\pi }{2N})^{T}  \label{6}
\end{equation}

, where $0\leq u<N$

The $N\times N$ DCT matrix can be expressed as

\begin{equation}
D=(\alpha _{u}\cos \frac{(2n+1)u\pi }{2N})_{N\times N}=\left[
\begin{array}{c}
(\overset{\rightarrow }{D_{0}})^{T} \\
(\overset{\rightarrow }{D_{1}})^{T} \\
\vdots \\
(\overset{\rightarrow }{D_{N-1}})^{T}%
\end{array}%
\right]  \label{7}
\end{equation}

Equation (4) can be expressed as

\begin{equation}
\overset{\rightarrow }{c}=\left(
\begin{array}{c}
c_{0} \\
c_{1} \\
\vdots \\
c_{N-1}%
\end{array}%
\right) =D\overset{\rightarrow }{f}=\left(
\begin{array}{c}
\overset{\rightarrow }{D_{0}}\bullet \overset{\rightarrow }{f} \\
\overset{\rightarrow }{D_{1}}\bullet \overset{\rightarrow }{f} \\
\vdots \\
\overrightarrow{D_{N-1}}\bullet \overset{\rightarrow }{f}%
\end{array}%
\right)  \label{8}
\end{equation}

where \textquotedblleft $\bullet $\textquotedblright\ expresses inner
product between two vectors.\newline

\subsubsection{Two-Dimensional Discrete Cosine Transform \ (2-D DCT)}

The definition of the two-dimensional DCT \cite{1,2} (2-D DCT ) for input
image $F$ and coefficients matrix $C$ (output) is

\begin{equation}
\begin{tabular}{c}
$\left\{
\begin{tabular}{c}
$c_{pq}=\alpha _{p}\alpha
_{q}\sum\limits_{m=0}^{N-1}\sum\limits_{n=0}^{N-1}f_{mn}\cos \frac{\pi
(2m+1)p}{2N}\cos \frac{\pi (2n+1)q}{2N}$ \\
$f_{mn}=\sum\limits_{m=0}^{N-1}\sum\limits_{n=0}^{N-1}\alpha _{p}\alpha
_{q}c_{pq}\cos \frac{\pi (2m+1)p}{2N}\cos \frac{\pi (2n+1)q}{2N}$%
\end{tabular}%
\right. $ \\
$(0\leq p,q,m,n<N)$%
\end{tabular}
\label{9}
\end{equation}

2-D DCT is closely related to the discrete Fourier transform. It is a
separable linear transformation. That is, the result of 2-D DCT may be
obtained by first taking transforms along the columns of $F$ and then along
the rows of that result \cite{1,2}. That is \cite{1,2},

\begin{equation}
C=DFD  \label{10}
\end{equation}

In image compression, the input image is divided into 8-by-8 or 16-by-16
blocks, and the 2-D DCT is computed for each block. The DCT coefficients are
then quantized, coded, and transmitted. The receiver decodes the quantized
DCT coefficients, computes the inverse 2-D DCT of each block, and then puts
the blocks back together into a single image.

Almost all of digital films such as VCD and digital pictures such as JPEG
files are compressed by using DCT currently. Real-time compressing and high
compression ratio are the main research topics of image compression \cite{2}%
. The time complexity of classical 1-D DCT and 2-D DCT is $O(N\log _{2}N)$
and $O(N^{2}\log _{2}N)$ respectively \cite{1,2}. The more $N$ is big, the
more the compression ratio is high if the whole input data sequence is
Markov chain and the order of chain is bigger than $N$, but the running time
increases drastically so that real-time compressing is impossible. That is,
finding a fast algorithm for large $N$ is significant.

\subsubsection{Two Properties of DCT of Image}

\begin{itemize}
\item \textbf{Property 1:} 2-DCT of image is an energy conservation
transform \cite{1,2}, i.e.,%
\begin{equation*}
\sum\limits_{m=0}^{N-1}\sum\limits_{n=0}^{N-1}(f_{mn})^{2}=\sum%
\limits_{p=0}^{N-1}\sum\limits_{q=0}^{N-1}(c_{pq})^{2}
\end{equation*}
\end{itemize}

1-D DCT is same too.

The property is utilized as halt criterion of algorithm in this paper. If we
find some DCT coefficients (e.g., $c_{00},c_{01},c_{10}$) such that $%
\sum\limits_{m=0}^{N-1}\sum\limits_{n=0}^{N-1}(f_{mn})^{2}\approx
(c_{00})^{2}+(c_{01})^{2}+(c_{10})^{2}$, we will halt algorithm. That is, $%
c_{00},c_{01},c_{10}$ keep the information of image approximately.

\begin{itemize}
\item \textbf{Property 2:} For typical images, many of the DCT coefficients
have values close to zero; these coefficients can be discarded without
seriously affecting the quality of the reconstructed image \cite{1,2}.
\end{itemize}

Fig. 2, Fig. 3 and Table 1 illustrate the property 2 \cite{3}.\ The property
is utilized to design quantum algorithm in this paper.

\bigskip

\begin{tabular}{c}
\begin{tabular}{c}
\begin{tabular}{cc}
{\Large Fig2 } & {\Large \ Fig3 } \\
{\small (suggested minification: 60\%)} & {\small (suggested minification:
60\%)}%
\end{tabular}
\\
{\Large Table 1}%
\end{tabular}%
\end{tabular}

\bigskip

\subsection{The Classical Parallel Circuit for Computing Inner Product}

\begin{notation}
$t_{A}$ denotes the running time of addition $a+b$ or subtration $a-b$.
\end{notation}

\begin{notation}
$t_{M}$ denotes the running time of multiplication $a\times b$.
\end{notation}

\begin{notation}
$t_{D}$ denotes the running time of division $\frac{a}{b}$. In general, the
time of division is little bigger than the time of multiplication.
\end{notation}

\begin{notation}
$t_{C}$ denotes the running time of comparison $a\leq b$ or $a\geq b$.
\end{notation}

In classical computer, there exist constants $m_{1}$ and $m_{2}$ such that $%
t_{M}\geq m_{1}t_{A}$ and $t_{D}\geq m_{2}t_{M}$. The values of $m_{1}$ and $%
m_{2}$ relate with the material architecture of central processing unit
(CPU). Different CPU has different values of $m_{1}$ and $m_{2}$ \cite{4}.
For the convenience of computation, without loss of generality, we suppose
\begin{equation*}
\begin{tabular}{cccc}
$t_{M}\geqslant 25t_{A}$, & $t_{M}=2t_{M}$ & and & $t_{A}\gg t_{C}$%
\end{tabular}%
\end{equation*}

The assumption does not affect the time complexity because time compelxity
is only associated with the order of parameter $N$. E.g., $O(\frac{\pi }{4}%
\sqrt{N})=O(\sqrt{N})$ and $O(\frac{\pi }{4}\sqrt{N})\neq O(N)$.

Let $N-demensional$ vector%
\begin{eqnarray*}
\overset{\rightarrow }{x} &=&(x_{0},x_{1},\cdots ,x_{N-1})^{T} \\
\overset{\rightarrow }{y} &=&(y_{0},y_{1},\cdots ,y_{N-1})^{T}
\end{eqnarray*}

The inner product is defined as%
\begin{equation*}
\overset{\rightarrow }{x}\bullet \overset{\rightarrow }{y}=\underset{k=0}{%
\overset{N-1}{\sum }}x_{k}\times y_{k}
\end{equation*}

The classical parallel circuit for computing inner product can be designed
easily. Fig. 4 illustrates a realization of the classical parallel circuit.

\bigskip

\begin{tabular}{c}
{\Large Insert Fig4 Here}%
\end{tabular}

\bigskip

\begin{notation}
$t_{I}$ denotes the running time of parallel computation of inner product.
\end{notation}

Clearly, $t_{I}=1\times t_{M}+\lceil \log _{2}N\rceil t_{A}$

when $N\leq 2^{4\times 25}$, $t_{I}\leq 1\times t_{M}+4\times (25t_{A})\leq
5t_{M}$

Because $N=2^{100}$ is a very giant number, we regard $t_{I}$ as%
\begin{equation}
t_{I}\leq 5t_{M}  \label{11}
\end{equation}

\subsection{Introduction of Quantum Computation}

Recent research shows that realizing image compression by using quantum
computer is possilbe. For example, Pang\ recently presents a quantum
algorithm to realize the encoding of Vector Quantization, which is very
faster than classical encoding algorithm \cite{5}. This paper presents a
qunatum algorithm to realize DCT for image compression.

\subsubsection{Introduction of Quantum Bit (qubit)}

Just as classical bit has state - either 0 or 1 -- a qubit also has a state.
Two possible states for a qubit are the states ${|0\rangle }$ and ${%
|1\rangle }$, which as you might guess correspond to the states $0$ and $1$
for a classical bit. Notation like `${|}$ ${\rangle }$' is called the Dirac
notation, and we'll be seeing it often, as it's the standard notation for
states in quantum mechanics. The difference between bits and qubit can be in
a state \emph{other} than ${|0\rangle }$ or ${|1\rangle }$. It is also
possible to form linear combinations of states, often called superpositions%
\begin{equation*}
{|\psi \rangle }=\alpha {|0\rangle }+\beta {|1\rangle }
\end{equation*}

The numbers $\alpha $ and $\beta $ are complex numbers. Put another way, the
state of a qubit is a vector in a two-dimensional complex vector space. The
special states ${|0\rangle }$ and ${|1\rangle }$ are known as computational
basis states, and form an orthonormal basis for this vector space.

We can examine a bit to determine whether it is in the state 0 or 1 in
classical computer. By contrast, when we measure a qubit, we get either the
result 0, with probability $|\alpha |^{2}$, or the result 1, with probability%
$|\beta |^{2}$. Naturally, $|\alpha |^{2}+|\beta |^{2}=1$. In general a
qubit is a unit vector in a two-dimensional complex vector space \cite{6}.

We explain the superpositions ${|\psi \rangle }=\alpha {|0\rangle }+\beta {%
|1\rangle }$ by analogy with sonic wave as following.

Suppose there are three persons Alice, Bob and you in a closed room. Alice
and Bob speak in a sample wave ${|A\rangle }=f_{A}(t)=e^{im\omega t}$ and ${%
|B\rangle }=f_{B}(t)=e^{in\omega t}$ respectively, where $m\neq n$ and they
are both integers. We can distinguish Alice from Bob because the two sample
waves are orthogonal (i.e., $\langle A|B\rangle =\int_{0}^{\frac{2\pi }{%
\omega }}e^{i(n-m)\omega t}dt=0$). When Alice speak in the closed room, your
ears will receive a sonic wave $g_{A}(t)=I_{A}e^{\phi _{A}}{|A\rangle }$,
where $I_{A}$ is the amplitude of the wave and the phase $\phi _{A}$ is
cause by the distance between Alice and you. If Alice and Bob speak
simultaneously, your ears will receive a superposition:%
\begin{equation*}
{|\psi _{AB}\rangle }=I_{A}e^{\phi _{A}}{|A\rangle }+I_{B}e^{\phi _{B}}{%
|B\rangle }
\end{equation*}

Let
\begin{equation*}
\begin{tabular}{c}
$I=\sqrt{(I_{A})^{2}+(I_{B})^{2}}$ \\
\begin{tabular}{ccc}
$\alpha _{A}=\frac{I_{A}}{I}e^{i\phi _{A}}$ & and & $\beta _{B}=\frac{I_{B}}{%
I}e^{i\phi _{B}}$%
\end{tabular}%
\end{tabular}%
\end{equation*}

Thus,%
\begin{equation*}
{|\psi _{AB}\rangle }=I(\alpha _{A}{|A\rangle }+\beta _{B}{|B\rangle })
\end{equation*}

Your ears can distinguish Alice's voice from the superposition ${|\psi
_{AB}\rangle }$. That is, '$+$' implies two sonic wave ${|A\rangle }$ and ${%
|B\rangle }$ exist in the superposition simultaneously and they can be
distinguished from the superposition. If Alice speaks very aloud (i.e., $%
|\alpha _{A}|^{2}\rightarrow 1$ or $|\beta _{B}|^{2}\rightarrow 0$) , you
will always hear Alice's voice. This case is analogous with the case $%
|\alpha |^{2}\rightarrow 1$ of quantum computation. If $|\alpha
|^{2}\rightarrow 1$, you will get the result 0 always, with probability $%
|\alpha |^{2}\approx 1$. This property is utilized to design quantum
algorithm such as Grover's algorithm. You can operate the two distinguished
sample wave ${|A\rangle }$ and ${|B\rangle }$ simultaneously. For example,
you can send the voice ${|\psi _{AB}\rangle }$ in a radio and change Alice's
volumes and Bob's volumes simultaneously by pushing the volume button on
radio. That is, performing once operation causes the changing of two sonic
waves simultaneously. This case is analogous with quantum parallelism. Fig.
5 shows the analogies between quantum superpositions and sonic wave.

\bigskip

\begin{tabular}{c}
{\Large Insert Fig5 Here}%
\end{tabular}

\bigskip

\subsubsection{Operation of Computation Acting on Qubits}

Classical computer circuits consist of wires and logic gates. The wires are
used to carry information around the circuit, while the logic gates perform
manipulations of the information, converting it from one form to another. It
is the fundamental of classical computation that classical computer circuits
can realize the operations of Boolean algebra. For example, classical NOT
gate makes 0 and 1 states interchanged. The operations of Boolean algebra
can also be realized on quantum computer by utilizing single qubit gates and
controlled-NOT gates. For example, quantum NOT gate takes the state ${|\psi
\rangle }=\alpha {|0\rangle }+\beta {|1\rangle }$ to the corresponding state
in which the role of ${|0\rangle }$ and ${|1\rangle }$ have interchanged.
All digital operation can be realized by utilizing unitary operation that is
the combination of some single qubit gates and controlled-NOT gates \cite{6}.

\subsubsection{Quantum Parallelism}

Quantum parallelism allows quantum computers to evaluate a function $f(x)$
for many different values of $x$ simultaneously. The power of quantum
computation is due to the fact that the state of a quantum computer can be a
superposition of basis states and we can perform an operation on multiple
quantum states simultaneously. For example, suppose $f(x):\{0,1\}\rightarrow
\{0,1\}$ is a function with a one-bit domain and range. We need at least two
times calculating for obtain the values $f(0)$ and $f(1)$ on classical
computer. For arbitrary function $f(x)$, there is quantum circuit $U_{f}$
that can transform ${|0,y\rangle }$ and ${|1,y\rangle }$ into ${|0,y\oplus
f(0)\rangle }$ and ${|1,y\oplus f(1)\rangle }$ respectively by performing
only one time calculating, where $\oplus $ indicates addition modulo 2. That
is, $\frac{\left\vert 0\right\rangle \left\vert y\right\rangle +\left\vert
1\right\rangle \left\vert y\right\rangle }{\sqrt{2}}\overset{U_{f}}{%
\longrightarrow }\frac{\left\vert 0\right\rangle \left\vert {y\oplus }%
f(0)\right\rangle +\left\vert 1\right\rangle \left\vert {y\oplus }%
f(1)\right\rangle }{\sqrt{2}}$, where '$+$' implies two states $\left\vert
0\right\rangle \left\vert y\right\rangle $ and $\left\vert 1\right\rangle
\left\vert y\right\rangle $ exist in the superposition of states
simultaneously. The formula implies also that the two states $\left\vert
0\right\rangle \left\vert y\right\rangle $ and $\left\vert 1\right\rangle
\left\vert y\right\rangle $ are converted to $\left\vert 0\right\rangle
\left\vert {y\oplus }f(0)\right\rangle $ and $\left\vert 1\right\rangle
\left\vert {y\oplus }f(1)\right\rangle $ simultaneously. That is, the values
$f(0)$ and $f(1)$ are calculated simultaneously \cite{6,7}.

\subsubsection{Introduction of quantum Fourier transform (QFT),
Klappenecker's DCT and Latorre's Quantum Representation of Image}

The quantum Fourier transform ($QFT$ \cite{6,7}) on an orthonormal basis $%
|0\rangle $, $|1\rangle $, $\cdots $, $|N-1\rangle $ is defined to be a
linear operator with the following action on the basis states,

\begin{equation*}
QFT|j\rangle =\frac{1}{\sqrt{N}}\overset{N-1}{\underset{k=0}{\sum }}e^{2\pi
ijk/N}|k\rangle
\end{equation*}

Equivalently, the action on an arbitrary state may be written

\begin{equation*}
QFT(\underset{j=0}{\overset{N-1}{\sum }}\alpha _{j}|j\rangle )=\underset{j=0}%
{\overset{N-1}{\sum }}\alpha _{j}(QFT|j\rangle )
\end{equation*}

Similar to other unitary operation, $QFT$ is unitary operation that only
acts on basis states. It's an error opinion that unitary operation can act
on coefficients of basis states. Figure 6 illustrates that $QFT$ only acts
on basis states $|0\rangle $ and $|1\rangle $.

\bigskip

\begin{tabular}{c}
{\Large Insert Fig6 Here}%
\end{tabular}

\bigskip

Klappenecker presents a method to realize DCT of types I, II, III, and IV on
a quantum computer by utilizing $QFT$ \cite{8}. \newline
Define DCT of types I as \cite{8}

\begin{equation*}
C_{N}^{I}:=\sqrt{(\frac{2}{N})}[\alpha _{i}\cos \frac{ij\pi }{N}]_{i,j=0...N}
\end{equation*}

The discrete sine transforms (DST) of types I denoted by $S_{N}^{I}$ is
defined accordingly. \cite{8}

Let discrete Fourier transform (DFT) be \cite{8}

\begin{equation*}
F_{N}=\frac{1}{\sqrt{N}}[\overset{N-1}{\underset{k=0}{\sum }}(e^{2\pi
ijk/N})]_{k,l=0..N-1}
\end{equation*}%
Let \cite{8}

\begin{equation*}
T_{N}=\left(
\begin{array}{cccccccc}
1 &  &  &  &  &  &  &  \\
& \frac{1}{\sqrt{2}} &  &  &  & \frac{i}{\sqrt{2}} &  &  \\
&  & \ddots &  &  &  & \ddots &  \\
&  &  & \frac{1}{\sqrt{2}} &  &  &  & \frac{i}{\sqrt{2}} \\
&  &  &  & 1 &  &  &  \\
&  &  & \frac{1}{\sqrt{2}} &  &  &  & -\frac{i}{\sqrt{2}} \\
&  & \ddots &  &  &  & \ddots &  \\
& \frac{1}{\sqrt{2}} &  &  &  & -\frac{i}{\sqrt{2}} &  &
\end{array}%
\right)
\end{equation*}

Klappenecker's DCT derives from QFT and depends on QFT. Indeed, the DST and
DCT can be recovered from the DFT by a base change

\begin{equation*}
T_{N}^{+}\centerdot F_{2N}\centerdot T_{N}=C_{N}^{I}\oplus iS_{N}^{I}
\end{equation*}

Since efficient quantum circuit for the DFT (i.e., QFT) are known, it
remains to find an efficient implementation of the matrix $T_{N}$. A quantum
circuit is proposed by Klappenecker to realize the matrix $T_{N}$. This is
the primitive idea of Klappenecker's DCT \cite{8}.

The result of QFT or Klappenecker's DCT seems to indicate that quantum
computer can be used to very quickly compute the Fourier transform, which
would be fantastically useful in a wide range of applications. However, that
is \textit{not} exactly the case; the Fourier transform or Klappenecker's
DCT is being performed on the information `hidden' in the amplitudes of the
quantum state. This information is not directly accessible to measurement.
The catch, of course, is that if the output state is measured, it will
collapse each qubit into the state $|0\rangle $ or $|1\rangle $, preventing
us from learning the transform result directly\cite{6}. In addition, the
contents of section 2.1 in this paper shows that Klappenecker's DCT cannot
be applied to DCT of image compression too. Klappenecker's DCT is useful on
many other signal processing maybe.

Latorre presentes a novel quantum representation for image compression \cite%
{19}. The entropy of images is, in general, very large. An image with large
entropy is hard to compress. The idea in Latorre's paper is to keep the
largest eigenvalues of the Schmidt decompositions when the picture is
written in a renormalization group manner, that is, the largest contribution
to the entropy in that basis. Latorre's algorithm is nice but it is not
competitive with jpeg. The reason is that jpeg uses details of how human
see. The quantization table used in jpeg is tailored to the human eye.
Unless there are quantum quantization methods are incorporated in Latorre's
algorithm, there is no way it is as efficient as jpeg.

\subsection{Introduction of Grover's Algorithm and Boyer's Algorithm}

The algorithm by Grover solves the problem of searching for an element with
a unique index $x_{0}$ in a list of $N$ unsorted elements, similar to
searching a database like a telephone directory when we know the number but
not the person's name \cite{7}. Grover's quantum searching algorithm \cite{9}
takes advantage of quantum mechanical properties to perform the search with
an efficiency of order $O(\sqrt{N})$ \cite{6}.

To implement the quantum search we need to construct a unitary operation
that discriminates between the marked item $x_{0}$ and the rest. The
following function,%
\begin{equation*}
f(x)=\left\{
\begin{tabular}{ccc}
$0$ & $if$ & $x\neq x_{0}$ \\
$1$ & $if$ & $x=x_{0}$%
\end{tabular}%
\right.
\end{equation*}

and its corresponding unitary operation (i.e., black box or oracle),%
\begin{equation*}
O|x\rangle =(-1)^{f(x)}|x\rangle
\end{equation*}

will do the job.

The Grover iteration is defined as \cite{5}

\begin{equation}
G=(2|\Psi \rangle \langle \Psi |-I)O  \label{12}
\end{equation}

on $N-\dim ensional$ Hilbert space, where $|\Psi \rangle =\frac{1}{\sqrt{N}}%
\sum\limits_{x=0}^{N-1}|x\rangle $.

The Grover iteration can be regarded as a rotation in the two-dimensional
space spanned by $|\Psi \rangle $ and$\ |x_{0}\rangle $. To see this it is
useful to let $|\alpha \rangle =\frac{1}{\sqrt{N-1}}\underset{x\neq x_{0}}{%
\sum |x\rangle }$ indicate a sum over all $x$ which are not solutions to the
search problem. Thus,
\begin{equation*}
|\Psi \rangle =\sqrt{\frac{N-1}{N}}|\alpha \rangle +\sqrt{\frac{1}{N}}%
|x_{0}\rangle
\end{equation*}%
so the initial state of the quantum computer is in the space spanned by $%
|x_{0}\rangle $ and $|\alpha \rangle $. Let%
\begin{equation*}
\sin \theta /2=\sqrt{1/N}
\end{equation*}

As Figure 7 shows, $G$ is a rotation in the two-dimensional space $%
span\{|x_{0}\rangle ,|\alpha \rangle \}$, rotating the state $|\Psi \rangle $
by $\theta $ radians per application of $G$. Repeated application of the
Grover iteration rotates the state vector close to $|x_{0}\rangle $. When
this is occurs, an observation in the computational basis produces with high
probability \cite{6}. This is the key of Grover's algorithm.

\bigskip

\begin{tabular}{c}
{\Large Insert Fig7 Here}%
\end{tabular}

\bigskip

Eli Biham's papers show that Grover's algorithm is suitable for the case of
an arbitrary initial amplitude distribution \cite{10, 11}. Gui-Lu Long
improves Grover's algorithm to 100\% successful probability even the number
is very small \cite{12, 13}.

Michel Boyer, et., al. present the following algorithm for the case that the
number of solutions $t$ is unknown \cite{14}.

\begin{algorithm}
\textbf{Boyer's algorithm}
\end{algorithm}

1. Initialize $m=1$ and set $\lambda =6/5$.

(Any value of $\lambda $ strictly between 1 and 4/3 would do.)

2. Choose $j$ uniformly at random among the nonnegative integers smaller
than m.

3. Apply $j$ iterations of Grover's algorithm starting from initial state $%
|\psi _{0}\rangle =\sum_{i}\frac{1}{\sqrt{N}}|i\rangle $

4. Observe the register: let $i$ be the outcome.

5. If $T[i]=x$, the problem is solved: exit.

6. Otherwise, set $m$ to $min(\lambda m,\sqrt{N})$ and go back to step 2.

Boyer's algorithm is derived from Grover's algorithm and has time complexity
$O(\sqrt{\frac{N}{t}})$. The case of $t=0$ is handled by it. It is a very
smart algorithm becuase it saves many quantum circuits comparing wiht the
other algorithms by using quantum counting technique that will needed
exponential order circuits \cite{6}. Boyer's algorithm is utilized in this
paper. The one of task of this paper is to design appropriate quantum
iteration (named quantum DCT iteration in this paper) according to DCT
properties to substitute for the Grover iteration in Boyer's algorithm.

In this paper, Each ket represents a register rather than a single qubit. In
this paper, similar to other quantum algorithm \cite{6,7}, ancilla qubits
are ignored.

\section{The Representation of Image by Using Quantum States}

\subsection{Classical Data Structure of 1-D DCT}

For a given vector $x=(x_{0},x1,\cdots ,x_{N-1})$, we can declare a BYTE
array "$BYTE$ $x[N]$" to store it by using c language, where c language is
compiler language of classical computer \cite{15}:

\begin{equation*}
x[0]=x_{0},x[1]=x_{_{1}},...,x[N-1]=x_{N-1}
\end{equation*}

There is a logical mapping to associate subscript with component of vector $%
x $:

\begin{equation}
\begin{tabular}{c}
$Mapping:i\longmapsto x[i]$ \\
($0\leq i<N$)%
\end{tabular}
\label{13}
\end{equation}

The logical mapping is necessary because it associate data with the
corresponding logical address. CPU accesses value $x[i]$ according to the
subscript $i$ (i.e., logical address).

The mapping is done by memory-management unit (MMU) of Operating System \cite%
{16}. The operation of access data is a very very fast operation so that the
time of access can be ignored when designing algorithm.

Fig8 illustrates the logical mapping. Fig9 illustrates the physical
realization of the logical mapping \cite{16}.

\bigskip

\begin{tabular}{c}
{\Large Insert Fig8 Here}%
\end{tabular}

\bigskip

\bigskip

\begin{tabular}{c}
{\Large Insert Fig9 Here}%
\end{tabular}

\bigskip

Similar to vector $x$, the vectors $\overset{\rightarrow }{f_{i}}$ (equation
2), $\overset{\rightarrow }{c}$ (equation 8), $\overrightarrow{D_{u}}$\
(equation 6), matrix $D$ (equation 7) and matrix $F$ (equation 1) can be
stored in array respectively, and the Operating System of classical computer
will establish the mapping (equation 13) automatically \cite{16}.

For example, we declared a two dimensional array "$BYTE$ $arrayImage[N][N]$"
to store matrix $F$. The mapping between position $(i,j)$ and pixel value $%
f_{ij}$ is defined as
\begin{equation}
Mapping:(i,j)\mapsto arrayImage[i][j]=f_{ij}  \label{14}
\end{equation}

The above mapping (equations 13 and 14) should be also kept in quantum
computation so that arbitrary component of vector or matrix is associated
with the corresponding subscript.

By the definition of DCT (equations (4) and (9)), $QFT$ and Klappenecker's
DCT cannot both keep the mapping. Therefore, More suitable quantum data
structure is required in order to keep the mapping and harness the power of
quantum computation for image compression.

\subsection{The Quantum Representation of Image}

\subsubsection{Data Structure of Quantum Representation of Image}

To keep the mapping in equations (13) and (14), the following database
technique is presented to represent image $F$ (equation (1)) in this paper:

First, all vectors $\overset{\rightarrow }{f_{i}}%
=(f_{0i},f_{1i},...,f_{(N-1)i})^{T}$ ($0\leq i<N$) are stored in a database.
Each vector is regard as a record with unique index $i$.

Second, all vectors are loaded into CPU simultaneously and form the
superposition of states $\frac{1}{\sqrt{N}}|ancilla1\rangle (\overset{N-1}{%
\underset{i=0}{\sum }}|i\rangle |\overset{\rightarrow }{f_{i}}\rangle
)|ancilla2\rangle $ by using quantum addressing scheme and unitary operation
LOAD.

LOAD operation that is denoted by $U_{L}$ is defined as%
\begin{equation*}
|i\rangle |0\rangle \cdots |0\rangle \overset{U_{L}}{\rightarrow }|i\rangle
|0\oplus f_{0i}\rangle \cdots |0\oplus f_{(N-1)i}\rangle
\end{equation*}

, where $\oplus $ denotes addition modulo 2, that can be realized by
utilizing controlled NOT operation \cite{6}.

In vector notation, \
\begin{equation}
|i\rangle |0\rangle \overset{U_{L}}{\mapsto }|i\rangle |0\oplus \overset{%
\rightarrow }{f_{i}}\rangle  \label{15}
\end{equation}

LOAD operation is the basic operation of quantum computer (\cite{6}, chapter
6).

Figure 10 illustrates the representation of image by using quantum states.

\bigskip

\begin{tabular}{c}
{\Large Insert Fig10 Here}%
\end{tabular}

\bigskip

The proposed representation of image in this paper keep the mapping in
equation (14) so that subscript$(j,i)$ is associated with corresponding
pixel value $f_{ji}$. State $\frac{1}{\sqrt{N}}|ancilla1\rangle (\overset{N-1%
}{\underset{i=0}{\sum }}|i\rangle |\overset{\rightarrow }{f_{i}}\rangle
)|ancilla2\rangle $ is entanglement state when $ancilla1$ and $ancilla2$ are
constants. Therefore, if we obtain value $i$ from the second register, we
will get the unique mapping vector $\overset{\rightarrow }{f_{i}}$ in third
register. Thus, the mapping is kept.

\subsubsection{The Time of Load Operation $U_{L}$}

\begin{notation}
$t_{L}$ denotes the time of performing one time operation $|i\rangle
|0\rangle \mapsto |i\rangle |0\oplus f_{ji}\rangle $, where $f_{ji}$ is a
component of vector $\overset{\rightarrow }{f_{i}}$.
\end{notation}

\begin{notation}
$Time(U)$ denotes the time of performing one time operation $U$.
\end{notation}

The time of LOAD operation $U_{L}$ in equation (15) is denoted by $%
Time(U_{L})$.

The time of data access (i.e., loading data into registers from memory) is
ignored in classical algorithm when design algorithm. It is clear that the
most efficient possible algorithm is in this model of computation (\cite{6},
chapter 6, section 6.5 or \cite{16}). Operation $|i\rangle |0\rangle \mapsto
|i\rangle |0\oplus f_{ji}\rangle $ is CNOT operation, and it is a very very
fast operation so that it's running time can be ignored when analyze the
time complexity of algorithm (\cite{6}, chapter 6). Because each component $%
f_{ji}$ is saved in memory independently, each component can be load into
register simultaneously by utilizing parallel circuit in classical case.
Furthermore, the classical circuit can be translated into a quantum
reversible circuit. \cite{6} Therefore, in general,
\begin{equation}
Time(U_{L})=t_{L}\approx 0  \label{16}
\end{equation}

\section{Quantum 1-D DCT Iteration}

\subsection{The Design of Oracle $O_{inner}$, $O_{inner}^{\prime }$ and $%
O_{f}$}

All of vectors$\overset{\rightarrow }{\text{ }D_{i}}$ in equations (5) and
(6) are stored in a database (Fig10). Each vector $\overset{\rightarrow }{%
D_{i}}$ has a unique index $i$, where $0\leq i<N$.

An arbitrary vectors $\overset{\rightarrow }{D_{i}}$ is loaded into
registers according to the corresponding index $i$ by using unitary
operation $U_{L}$ (Fig10):

\begin{equation}
|\alpha \rangle |\beta \rangle |i\rangle |\overset{\rightarrow }{f}\rangle
|0\rangle |0\rangle |0\rangle \overset{U_{L}}{\rightarrow }|\alpha \rangle
|\beta \rangle |i\rangle |\overset{\rightarrow }{f}\rangle |\overset{%
\rightarrow }{D_{i}}\rangle |0\rangle |0\rangle  \label{17}
\end{equation}

where $\alpha $ and $\beta $ is two input parameters.

\subsubsection{Oracle $O_{inner}$ and $O_{inner}^{\prime }$}

It's easy to construct a classical parallel circuit to calculate inner
product (Section I-C and Fig. 4). Furthermore, the classical circuit can be
translated into a quantum reversible circuit \cite{6}, that has the same
time complexity with it's corresponding classical circuit.

We design oracles $O_{inner}$ and $O_{inner}^{\prime }$to compute the inner
product between two vectors$\overset{\rightarrow }{\text{ }D_{i}}$ and $%
\overset{\rightarrow }{f}$.

\begin{equation}
|\alpha \rangle |\beta \rangle |i\rangle |\overset{\rightarrow }{f}\rangle |%
\overset{\rightarrow }{D_{i}}\rangle |0\rangle |0\rangle \overset{O_{inner}}{%
\rightarrow }|\alpha \rangle |\beta \rangle |i\rangle |\overset{\rightarrow }%
{f}\rangle |\overset{\rightarrow }{D_{i}}\rangle |0\rangle |(\overset{%
\rightarrow }{D_{i}}\bullet \overset{\rightarrow }{f})^{2}\rangle  \label{18}
\end{equation}

\begin{equation}
|\alpha \rangle |\beta \rangle |i\rangle |\overset{\rightarrow }{f}\rangle |%
\overset{\rightarrow }{D_{i}}\rangle |0\rangle |0\rangle \overset{%
O_{inner}^{\prime }}{\rightarrow }|\alpha \rangle |\beta \rangle |i\rangle |%
\overset{\rightarrow }{f}\rangle |\overset{\rightarrow }{D_{i}}\rangle |%
\overset{\rightarrow }{D_{i}}\bullet \overset{\rightarrow }{f}\rangle
|0\rangle  \label{19}
\end{equation}

Fiure 11 illustrates the oracles $O_{inner}$.

\bigskip

\begin{tabular}{c}
{\Large Insert Fig11 Here}%
\end{tabular}

\bigskip

By equation (11), oracles $O_{inner}$ has time complexity%
\begin{equation}
Time(O_{inner})=t_{I}+1\times t_{M}\leq 6t_{M}  \label{20}
\end{equation}

Oracles $O_{inner}^{\prime }$ has time complexity
\begin{equation}
Time(O_{inner}^{\prime })=t_{I}\leq 5t_{M}  \label{21}
\end{equation}

\subsubsection{Oracle $O_{f}$}

Define a function $f$ as following%
\begin{equation*}
f(i)=\left\{
\begin{tabular}{ccc}
1 & if & $\alpha \leq (\overset{\rightarrow }{D_{i}}\bullet \overset{%
\rightarrow }{f})^{2}\leq \beta $ \\
0 & \multicolumn{2}{c}{otherwise}%
\end{tabular}%
\right.
\end{equation*}

We design other oracle denoted by $O_{f}$ that is used to mark the retained
DCT coefficients (Table 1):

\begin{multline}
|\alpha \rangle |\beta \rangle |i\rangle |\overset{\rightarrow }{f}\rangle |%
\overset{\rightarrow }{D_{i}}\rangle |0\rangle |(\overset{\rightarrow }{D_{i}%
}\bullet \overset{\rightarrow }{f})^{2}\rangle \overset{O_{f}}{\rightarrow }
\label{22} \\
(-1)^{f(i)}|\alpha \rangle |\beta \rangle |i\rangle |\overset{\rightarrow }{f%
}\rangle |\overset{\rightarrow }{D_{i}}\rangle |0\rangle |(\overset{%
\rightarrow }{D_{i}}\bullet \overset{\rightarrow }{f})^{2}\rangle
\end{multline}

Figure 12 illustrates the oracles $O_{f}$.

\bigskip

\begin{tabular}{c}
{\Large Insert Fig12 Here}%
\end{tabular}

\bigskip

Oracle $O_{f}$ has time complexity .%
\begin{equation}
Time(O_{f})=2t_{C}  \label{23}
\end{equation}

\subsection{Quantum 1-D DCT Iteration $G_{DCT}$}

Let $(O_{inner})^{-1}$\ denote the inverse transform of $O_{inner}$.

Let $(U_{L})^{-1}$ denote the inverse transform of $U_{L}$.%
\begin{equation*}
(O_{inner}U_{L})^{-1}=(U_{L})^{-1}(O_{inner})^{-1}
\end{equation*}

\begin{definition}
By\ equations (17),\ (18)\ and\ (22),\ \emph{quantum\ 1-D\ DCT\ iteration}\ $%
G_{DCT}$\ is defined as
\end{definition}

\begin{equation}
G_{DCT}=(2|\xi \rangle \langle \xi
|-I)(U_{L})^{-1}(O_{inner})^{-1}O_{f}O_{inner}U_{L}  \label{24}
\end{equation}%
, where $|\xi \rangle =\frac{1}{\sqrt{N}}\sum\limits_{i=0}^{N-1}|i\rangle $.

We have%
\begin{multline}
(U_{L})^{-1}(O_{inner})^{-1}O_{f}O_{inner}U_{L}|\alpha \rangle |\beta
\rangle |i\rangle |\overset{\rightarrow }{f}\rangle |0\rangle |0\rangle
|0\rangle =  \label{25} \\
(-1)^{f(i)}|\alpha \rangle |\beta \rangle |i\rangle |\overset{\rightarrow }{f%
}\rangle |0\rangle |0\rangle |0\rangle
\end{multline}

Equation (25) shows that the state $|\alpha \rangle |\beta \rangle |i\rangle
|\overset{\rightarrow }{f}\rangle |0\rangle |0\rangle |0\rangle $ is the
eigenstate of unitary operation $%
(U_{L})^{-1}(O_{inner})^{-1}O_{f}O_{inner}U_{L}$ and the eigenvalue is $%
(-1)^{f(i)}$. This is a key of applying two or more oracles to perform
complex search.

Figure 13 shows the action of unitary $%
(U_{L})^{-1}(O_{inner})^{-1}O_{f}O_{inner}U_{L}$.

\bigskip

\begin{tabular}{c}
{\Large Insert Fig13 Here}%
\end{tabular}

\bigskip

Let $span\{|i\rangle \otimes |vector\rangle \otimes |innerproduct\rangle \}$
be the global space that spanned by $|i\rangle $, $N-dimensional$ vector $%
|vector\rangle $, and the value of inner product $|innerproduct\rangle $.

Let $span\{|i\rangle \mid 0\leq i<N\}$ be the subspace that is the span of
all states $|i\rangle .$

\begin{theorem}
(Figure 14, 15) $G_{DCT}$ is a rotation on $N-\dim ensional$ subspace $%
span\{|i\rangle \mid 0\leq i<N\}$. For initial state $|\psi _{0}\rangle =%
\frac{1}{\sqrt{N}}\sum\limits_{i=0}^{N-1}(|\alpha \rangle |\beta \rangle
|i\rangle |\overset{\rightarrow }{f}\rangle |0\rangle |0\rangle |0\rangle )$%
, the rotation angle is $\theta =2\arcsin \sqrt{\frac{t}{N}}$, where $t$ is
the number of solutions.
\end{theorem}

\begin{proof}
For arbitrary input state
\begin{equation*}
|input\rangle =\sum\limits_{i=0}^{N-1}A_{i}(|\alpha \rangle |\beta \rangle
|i\rangle |\overset{\rightarrow }{f}\rangle |0\rangle |0\rangle |0\rangle )
\end{equation*}

where $A_{i}$ is the probability amplitude of the state and%
\begin{equation*}
\sum\limits_{i=0}^{N-1}\mid A_{i}\mid ^{2}=1
\end{equation*}

By equation (25), we have%
\begin{multline}
G_{DCT}|input\rangle =|\alpha \rangle |\beta \rangle  \label{26} \\
\{\underbrace{(2|\xi \rangle \langle \xi
|-I)(\sum\limits_{i=0}^{N-1}(-1)^{f(i)}A_{i}|i\rangle )}\}|\overset{%
\rightarrow }{f}\rangle |0\rangle |0\rangle |0\rangle
\end{multline}

Equation (26) shows that unitary operation $2|\xi \rangle \langle \xi |-I$
only acts on state $\sum\limits_{i=0}^{N-1}(-1)^{f(i)}A_{i}|i\rangle $.

Suppose that the solutions set is
\begin{equation*}
S=\{|\alpha \rangle |\beta \rangle |i\rangle |\overset{\rightarrow }{f}%
\rangle |0\rangle |0\rangle |0\rangle \mid \alpha \leq (\overset{\rightarrow
}{D_{i}}\bullet \overset{\rightarrow }{f})^{2}\leq \beta \}
\end{equation*}

All solutions form a superposition%
\begin{equation*}
|\tau \rangle =\frac{1}{\sqrt{|S|}}\underset{|u\rangle \in S}{\sum }%
|u\rangle =\frac{1}{\sqrt{|S|}}|\alpha \rangle |\beta \rangle \underbrace{%
(\sum_{S}|i\rangle )}|\overset{\rightarrow }{f}\rangle |0\rangle |0\rangle
|0\rangle
\end{equation*}

Let
\begin{multline*}
|\sigma \rangle =\frac{1}{\sqrt{N-|S|}}\underset{|v\rangle \notin S}{\sum }%
|v\rangle = \\
\frac{1}{\sqrt{N-|S|}}|\alpha \rangle |\beta \rangle \underbrace{%
(\sum_{I-S}|j\rangle )}|\overset{\rightarrow }{f}\rangle |0\rangle |0\rangle
|0\rangle
\end{multline*}

The initial state $|\psi _{0}\rangle $ is rotated in the subspace $%
span\{|\sigma \rangle $,$|\tau \rangle \}$ by $\theta $\ towards the
superposition $|\tau \rangle $\ of all solutions to the search.

Initially, it is inclined at angle $\frac{\theta }{2}$\ from $|\sigma
\rangle $, a state orthogonal to $|\tau \rangle $.

By equation (25), the product operation $%
(U_{L})^{-1}(O_{inner})^{-1}O_{f}O_{inner}U_{L}$ reflects the state about
the state $|\sigma \rangle $. Then the operation $2|\xi \rangle \langle \xi
|-I$ reflect it about $|\xi \rangle $.

Notice that $|\tau \rangle \in span\{|i\rangle \}$ and $|\sigma \rangle \in
span\{|i\rangle \}$.

Therefore, similar to Grover's iteration, the quantum 1-D DCT iteration $%
G_{DCT}$ is a rotation on subspace $span\{|i\rangle \}$, and the rotation
angle is $\theta =2\arcsin \sqrt{\frac{t}{N}}$.
\end{proof}

Fig14 illustrates the action of a single $G_{DCT}$.

Fig15 shows that $G_{DCT}$ is equivalent to a rotation on subspace $%
span\{|i\rangle \}$.

\bigskip

\begin{tabular}{c}
{\Large Insert Fig14 Here}%
\end{tabular}

\bigskip

\bigskip

\begin{tabular}{c}
{\Large Insert Fig15 Here}%
\end{tabular}

\bigskip

1-D DCT iteration $G_{DCT}$ has time complexity%
\begin{equation*}
Time(G_{DCT})=2Time(U_{L})+2Time(O_{inner})+Time(O_{f})
\end{equation*}

By equations (16), (20) and (23), when $N\leq 2^{100}$,

\begin{equation*}
Time(G_{DCT})\leq 2t_{C}+12t_{M}
\end{equation*}

Because $t_{M}\geq 25t_{A}\gg t_{C}$, we have%
\begin{equation}
Time(G_{DCT})\leq 13t_{M}  \label{27}
\end{equation}

The theorem shows that the quantum 1-D DCT iteration $G_{DCT}$ and the
Grover iteration $G=(2|\Psi \rangle \langle \Psi |-I)O$ have the same
property that they are both a rotation on space. However, comparing with
Grover iteration that has only one oracle, $G_{DCT}$ comprises two oracles $%
O_{inner}$ and $O_{f}$ so that it can perform complex search. Equation (25)
is a key of applying two or more oracles to perform complex search. If there
is no equation (25), applying two or more oracles is no significant.

\section{Quantum 1-D DCT of Image Compression}

\subsection{Subroutine 1: Finding a DCT Coefficient $c_{i_{0}}\in
\{c_{i}\mid \protect\alpha \leq (c_{i})^{2}\leq \protect\beta \}$}

Subroutine 1 is similar to Boyer's algorithm, and the Grover iteration in
Boyer's algorithm is substituted by the iteration $G_{DCT}$.

\begin{algorithm}
\textbf{Subroutine 1}
\end{algorithm}

\textbf{Input parameters:} $\alpha ,\beta $

\textbf{Outputs:} subscript $i_{0}$ and coefficient $c_{i_{0}}$; a global
Boolean variable $IsSolution$.

\textbf{Step1.} Initialize $m=1$ and $\lambda =\frac{6}{5}$; $%
IsSolution=FALSE$;

\textbf{Step2}. Choose $j$ uniformly at random among the nonnegative
integers not bigger than $m$.

\textbf{Step3}. Generate the initial state

\begin{equation*}
|\Psi _{0}\rangle =\frac{1}{\sqrt{N}}|\alpha \rangle |\beta \rangle
\underbrace{(\sum\limits_{i=0}^{N-1}|i\rangle )}|\overset{\rightarrow }{f}%
\rangle |0\rangle |0\rangle |0\rangle
\end{equation*}

\ \ \ \ \ This can be achieved in $O(log_{2}N)$ steps using a $\lceil \log
_{2}N\rceil -bit$ Hadamard transformation.

\textbf{Step4.} Apply $j$ iterations of $G_{DCT}$ starting from the state $%
|\Psi _{0}\rangle $. i.e.,%
\begin{equation*}
|\Psi _{j}\rangle =(G_{DCT})^{j}|\Psi _{0}\rangle
\end{equation*}

\textbf{Step5.} Perform computation .%
\begin{equation*}
|\Psi _{end}\rangle =O_{inner}^{\prime }U_{L}|\Psi _{j}\rangle
\end{equation*}

\textbf{Step6.} Observe the third and the fifth register: let $i_{0}$, $%
c_{i_{0}}$ be the outcome respectively.

\textbf{Step7.} If $c_{i_{0}}\in \{c_{i}\mid \alpha \leq (c_{i})^{2}\leq
\beta \}$, Let $IsSolution=TRUE$, and \emph{exit}; otherwise, go to step 8.

\textbf{Step8.} set $m$ to $\min \{\lambda m,\sqrt{N}\}$ and go to step 2.

Figure 16 shows the processing of subroutine 1.

\bigskip

\begin{tabular}{c}
{\Large Insert Fig16 Here }(suggested minification: 78\%)%
\end{tabular}

\bigskip

\subsection{The Correctness of Subroutine 1}

When $2^{8}\leq N\leq 2^{100}$, $|\frac{\sin (\frac{1}{\sqrt{N}})}{\frac{1}{%
\sqrt{N}}}-1|\leq 6.5\times 10^{-4}$, $99.6\%\leq 1-\frac{1}{N}<1$. That is,
when $2^{8}\leq N\leq 2^{100}$, Grover's algorithm is still an efficient
possible algorithm and the success probability $p=1-o(\frac{1}{N})\geq
99.6\% $ \cite{6}. Long improves Grover's algorithm so that the improved
algorithm with 1$00\%$ success probability even $N$ is very small \cite{12,
13}.

The theorem in this paper demonstrates that $G_{DCT}$ is similar to the
Grover iteration.

Therefore, subroutine 1 is correct.

\subsection{The Time Complexity of Subroutine 1}

Let $\kappa $ presents the iteration steps of applying $G_{DCT}$.

Subroutine 1 has time complexity%
\begin{multline*}
Time(subroutine1)\leq \kappa \lbrack Time(G_{DCT})+Time(O_{inner}^{\prime })
\\
+\lceil \log _{2}N\rceil Time(Hadamard)+2t_{m}+3t_{c}]
\end{multline*}

, where $Time(Hadamard)$ represents the time of Hadamard transform acting on
a single qubit.

when $N\leq 2^{100}$, we have
\begin{equation*}
\log _{2}NTime(Hadamard)\approx 0
\end{equation*}

Because $t_{M}\geq 25t_{A}\gg t_{C}$, by equations (21) and (27)%
\begin{equation*}
Time(subroutine1)\leq 21\kappa t_{M}
\end{equation*}

Subroutine 1 is similar to Boyer's algorithm. Hence
\begin{equation*}
\kappa =O(\sqrt{\frac{N}{t}})
\end{equation*}

Therefore,%
\begin{equation}
Time(subroutine1)=O(\sqrt{\frac{N}{t}})t_{M}  \label{28}
\end{equation}

\subsection{Quantum 1-D DCT of Image Compression}

\begin{algorithm}
\textbf{Quantum 1-D DCT of Image Compression} (\textbf{QDCT1})
\end{algorithm}

\textbf{Input: }$\overset{\rightarrow }{\{D}_{i}|0\leq i<N\}$: The line
vectors of DCT matrix (equation 5, 6, and 7). These line vectors are stored
in database with indices $i$ (Fig10);

Vector $\overset{\rightarrow }{f}$ (equation 2)

\textbf{Parameters:}$\Delta E$, $\alpha $, $\beta $, threshold value $%
\varepsilon >0$;

$nCount_{i}$ ($0\leq i<N$): Counting the repeated times of coefficient $%
c_{i} $ that is found when repeatedly applying subroutine 1.

$nMaxRepetition$: The maximum number of allowed repeated times of applying
subroutine 1.

$nS$: The number of solutions that have been obtained.

\textbf{Output: }The big DCT coefficients and corresponding indices $%
(i_{0},c_{i_{0}})$, $(i_{1},c_{i_{1}},)$, $\cdots $, $(i_{R},c_{i_{R}})$
such that $\frac{\overset{R}{\underset{k=0}{\sum }}(c_{i_{k}}\times
c_{i_{k}})}{||\overset{\rightarrow }{f}||^{2}}\leq \varepsilon $, where $||%
\overrightarrow{f}||^{2}=(\overrightarrow{f}\bullet \overrightarrow{f})=%
\underset{i=0}{\overset{N-1}{\sum }}(f_{i}\times f_{i})$. The other
coefficients are set to zero (Table 1).

\textbf{Step 0 (Initialize parameters):} $\Delta E=||\overrightarrow{f}%
||^{2} $, $\alpha =\frac{\Delta E}{N}$, $\beta =\Delta E$; $nS=0$; $%
nCount_{i}=0$, where $0\leq i<N$.

The proposed parallel circuit in section I-C (figure 4) can be applied to
calculate the value $||\overrightarrow{f}||^{2}=(\overrightarrow{f}\bullet
\overrightarrow{f})$.

while($\frac{\Delta E}{||\overset{\rightarrow }{f}||^{2}}\geq \varepsilon $)

\{

\ \ \ \textbf{Step 1:} Apply subroutine 1 to find a coefficient $%
(i_{k},c_{i_{k}})$;

\ \ \ \textbf{Step 2:} If $IsSolution=TRUE$ and $nCount_{i}=0$,

\ \ \ \ \ \ \ \ \ \ \ \ then $\Delta E=\Delta E-c_{i_{k}}\times c_{i_{k}}$; $%
nS=nS+1$; $\alpha =\frac{\Delta E}{N-nS}$, $\beta =\Delta E$; $%
nCount_{i_{k}}=nCount_{i_{k}}+1$;

\ \ \ \ \ \ \ \ \ \ If $IsSolution=TRUE$ and $%
0<nCount_{i_{k}}<nMaxRepetition $, continue;

\ \ \ \ \ \ \ \ \ \ If $IsSolution=TRUE$, and $nCount_{i_{k}}\geq
nMaxRepetition$, \textbf{exit} and apply classical algorithm to obtain DCT
coefficients;

\ \ \ \ \ \ \ \ \ \ If $IsSolution=FALSE$, \textbf{exit} and apply classical
algorithm to obtain DCT\ coefficients;

\}

\begin{example}
We choose the first 8 number gray value of the first line of image in fig.\
2 as vector $\overset{\rightarrow }{f}$. That is,
\end{example}

\begin{equation*}
\begin{tabular}{c}
$\overset{\rightarrow }{f}=(f_{00},f_{01},f_{02},\cdots ,f_{07})^{T}=$ \\
$(156,159,158,155,158,156,159,158)^{T}$%
\end{tabular}%
\end{equation*}

\begin{equation*}
\overset{\rightarrow }{c=(c_{0},c_{1},c_{2},\cdots ,c_{7})=}DCT(\overset{%
\rightarrow }{f})
\end{equation*}

\begin{equation*}
\begin{tabular}{c}
$((c_{0})^{2},(c_{1})^{2},(c_{2})^{2},\cdots ,(c_{7})^{2})=$ \\
${\small (1.9814e+005,\qquad 0.51531,\qquad 1.5063,\qquad 0.95824,}$ \\
${\small 3.125,\qquad 2.5846,\qquad 2.7437,\qquad 4.4418)}^{T}$%
\end{tabular}%
\end{equation*}

Let $\varepsilon =2.0\times 10^{-5}$

The computation\ procedure of 1-D QDCT is listed below:

\textbf{Step A}(Fig17). \textbf{Input:} $\Delta E=||\overset{\rightarrow }{f}%
||^{2}=198151$, $\beta =\Delta E$, $\alpha =\frac{\Delta E}{8}=24768.875$;

\textbf{Output:}\ The set of solutions $S=\{(k,c_{k})\mid \alpha \leq
(c_{k})^{2}\leq \beta \}=\{(0,c_{0})\}$. The output is the unique solution $%
(0,c_{0})$.

\bigskip

\begin{tabular}{c}
{\Large Insert Fig17 Here}%
\end{tabular}

\bigskip

\textbf{Step B }(Fig18). \textbf{Input:} $\Delta E=\Delta E-(c_{0})^{2}=11$,
$\frac{\Delta E}{||\overset{\rightarrow }{f}||^{2}}=5.55132\times 10^{-5}$, $%
\beta =\Delta E$, $\alpha =\frac{\Delta E}{8-1}=1.57143$;

\textbf{Output: }The set of solutions $S=%
\{(4,c_{4}),(5,c_{5}),(6,c_{6}),(7,c_{7})\}$.

We will obtain one of solution with equal probability. Suppose the output is
$(6,c_{6})$.

\bigskip

\begin{tabular}{c}
{\Large Insert Fig18 Here}%
\end{tabular}

\bigskip

\textbf{Step C}. \textbf{Input:} $\Delta E=\Delta E-(c_{6})^{2}=8.2563$, $%
\frac{\Delta E}{||\overset{\rightarrow }{f}||^{2}}=4.16667\times 10^{-5}$, $%
\beta =\Delta E$, $\alpha =\frac{\Delta E}{8-2}=1.37605$;

\textbf{Output: }The set of solutions $S=%
\{(2,c_{2}),(4,c_{4}),(5,c_{5}),(7,c_{7}),(6,c_{6})\}$.

We will obtain a new solution with probability $p=\frac{4}{5}$. That is,
repeating subroutine 1 $\lceil \frac{1}{p}\rceil $ times approximately, we
will obtain a new solution. Suppose the output is $(7,c_{7})$.

Because $\frac{\Delta E-(c_{7})^{2}}{||\overset{\rightarrow }{f}||^{2}}%
=1.92505\times 10^{-5}<\varepsilon $, stop.

\subsection{Time Complexity of Quantum 1-D DCT of Image Compression (QDCT1)}

Suppose the number of retained DCT coefficients is $M$.

For typical images, many of the DCT coefficients have values close to zero;
these coefficients can be discarded without seriously affecting the quality
of the reconstructed image \cite{3} (section I-B.3., Table 1).

The above example shows that the only one retained DCT coefficientthe is $%
c_{0}$ if $\varepsilon =5.56\times 10^{-5}$. That is, $c_{0}$ keeps almost
information of vector $\overset{\rightarrow }{f}.$ Thus, $M=1$ (i.e., $M\ll
N $)

By equation (11), the step0 has complexity
\begin{equation*}
Time(Step0)=t_{I}+t_{D}\leq 7t_{M}
\end{equation*}

The step1 has complexity%
\begin{equation*}
Time(Step1)=Time(subroutine1)
\end{equation*}

The step2 has complexity%
\begin{equation*}
Time(Step2)=t_{M}+t_{D}+3t_{A}+8t_{C}\leq 5t_{M}
\end{equation*}

Thus, the time complexity of QDCT1 is

\begin{multline*}
Time(QDCT1)=Time(Step0)+ \\
M(t_{D}+Time(Step1)+Time(Step2)) \\
\leq 7t_{M}+M(7t_{M}+Time(subroutine1))
\end{multline*}

Because $M\ll N$, by equation (28), we have

\begin{equation}
Time(QDCT1)=O(\sqrt{N})t_{M}  \label{29}
\end{equation}

\section{Quantum 2-D DCT}

By formula (10), the 2-D DCT can be expressed as%
\begin{equation}
C=DFD=(DF)D=GD  \label{30}
\end{equation}%
,where%
\begin{equation}
\begin{tabular}{c}
$G=DF=(g_{ij})_{N\times N}=(\overrightarrow{D_{i}}\bullet \overrightarrow{%
f_{j}})_{N\times N}=$ \\
$=\left(
\begin{array}{c}
\overset{\rightarrow }{D}_{0} \\
\overset{\rightarrow }{D}_{1} \\
\vdots \\
\overrightarrow{D_{N-1}}%
\end{array}%
\right) \left(
\begin{array}{cccc}
\overset{\rightarrow }{f}_{0} & \overset{\rightarrow }{f}_{1} & ... &
\overset{\rightarrow }{f}_{N-1}%
\end{array}%
\right) $%
\end{tabular}
\label{31}
\end{equation}

\subsection{The Data Structure of Computing Matrix $G$}

2-D DCT is closely related to the discrete Fourier transform. It is a
separable linear transformation, that is, the result of 2-D DCT may be
obtained by first taking transforms along the columns of $F$ and then along
the rows of that result \cite{1,2}.

Clearly, $G$\ is the result of 1-D DCT by taking transforms along the
columns of $F$.

All of vectors $\overset{\rightarrow }{D}_{i}$ and $\overset{\rightarrow }{f}%
_{j}$are stored in two different tables of database, where $0\leq i,j<N$.

An arbitrary record $\overset{\rightarrow }{D}_{i}$ or $\overset{\rightarrow
}{f}_{j}$can be fetched into registers of CPU according to indices $i,j$ by
performing load operation $U_{L}^{\prime }$.

Load operation $U_{L}^{\prime }$ is defined as%
\begin{equation}
|\alpha \rangle |\beta \rangle \underbrace{|i\rangle |j\rangle }|0\rangle
|0\rangle |0\rangle |0\rangle \overset{U_{L}^{\prime }}{\rightarrow }|\alpha
\rangle |\beta \rangle \underbrace{|i\rangle |j\rangle }|\overset{%
\rightarrow }{f_{j}}\rangle |\overset{\rightarrow }{D_{i}}\rangle |0\rangle
|0\rangle  \label{32}
\end{equation}

We design oracles $B_{inner}$ and $B_{inner}^{\prime }$ to compute the inner
product between two vectors $\overset{\rightarrow }{D_{i}}$ and $\overset{%
\rightarrow }{f}_{j}$%
\begin{multline}
|\alpha \rangle |\beta \rangle |i\rangle |j\rangle |\overset{\rightarrow }{%
f_{j}}\rangle |\overset{\rightarrow }{D_{i}}\rangle |0\rangle |0\rangle
\overset{B_{inner}}{\rightarrow } \\
|\alpha \rangle |\beta \rangle |i\rangle |j\rangle |\overset{\rightarrow }{%
f_{j}}\rangle |\overset{\rightarrow }{D_{i}}\rangle |0\rangle |(\overset{%
\rightarrow }{D_{i}}\bullet \overset{\rightarrow }{f_{j}})^{2}\rangle
\label{33}
\end{multline}

\begin{multline*}
|\alpha \rangle |\beta \rangle |i\rangle |j\rangle |\overset{\rightarrow }{%
f_{j}}\rangle |\overset{\rightarrow }{D_{i}}\rangle |0\rangle |0\rangle
\overset{B_{inner}^{\prime }}{\rightarrow } \\
|\alpha \rangle |\beta \rangle |i\rangle |j\rangle |\overset{\rightarrow }{%
f_{j}}\rangle |\overset{\rightarrow }{D_{i}}\rangle |(\overset{\rightarrow }{%
D_{i}}\bullet \overset{\rightarrow }{f_{j}})\rangle |0\rangle
\end{multline*}

Define a function $f^{\prime }$ as following%
\begin{equation*}
f^{\prime }(i,j)=\left\{
\begin{tabular}{ccc}
$1$ & $if$ & $\alpha \leq (\overset{\rightarrow }{D_{i}}\bullet \overset{%
\rightarrow }{f}_{j})^{2}\leq \beta $ \\
$0$ & \multicolumn{2}{c}{otherwise}%
\end{tabular}%
\right.
\end{equation*}

We design other oracle denoted by $O_{f}^{\prime }$ that is used to mark the
retained elements of $G$:

\begin{multline}
|\alpha \rangle |\beta \rangle |i\rangle |j\rangle |\overset{\rightarrow }{%
f_{j}}\rangle |\overset{\rightarrow }{D_{i}}\rangle |0\rangle |(\overset{%
\rightarrow }{D_{i}}\bullet \overset{\rightarrow }{f_{j}})^{2}\rangle
\overset{O_{f}^{\prime }}{\rightarrow } \\
(-1)^{f^{\prime }(i)}|\alpha \rangle |\beta \rangle |i\rangle |j\rangle |%
\overset{\rightarrow }{f_{j}}\rangle |\overset{\rightarrow }{D_{i}}\rangle
|0\rangle |(\overset{\rightarrow }{D_{i}}\bullet \overset{\rightarrow }{f_{j}%
})^{2}\rangle  \label{34}
\end{multline}

Let $(B_{inner})^{-1}$\ denote the inverse transform of $B_{inner}$.

Let $(U_{L}^{\prime })^{-1}$ denote inverse transform of $U_{L}^{\prime }$.

\begin{definition}
By equations (32), (33) and (34), \emph{quantum 2-D DCT iteration} $%
G_{DCT}^{\prime }$ is defined as
\end{definition}

\begin{equation}
G_{DCT}^{\prime }=(2|\xi ^{\prime }\rangle \langle \xi ^{\prime
}|-I)(U_{L}^{\prime })^{-1}(B_{inner})^{-1}O_{f}^{\prime
}B_{inner}U_{L}^{\prime }  \label{35}
\end{equation}%
, where$\mid \xi ^{\prime }>=\frac{1}{N}\sum\limits_{i=0}^{N-1}\sum%
\limits_{j=0}^{N-1}\underbrace{|i\rangle |j\rangle }$.

\subsection{Subroutine 2: Finding an Element $g_{i_{0}j_{0}}\in \{g_{ij}\mid
\protect\alpha \leq (g_{ij})^{2}\leq \protect\beta \}$}

Subroutine 2 is similar to Subroutine 1.

\begin{algorithm}
\textbf{Subroutine 2}
\end{algorithm}

\textbf{Input parameters:} $\alpha ,\beta $

\textbf{Outputs:} subscript $(i_{0},j_{0})$ and corresponding\ coefficient $%
g_{i_{0}j_{0}}$; \ a global Boolean variable $IsSolution$.

\textbf{Step1.} Initialize $m=1$ and $\lambda =\frac{6}{5}$; $%
IsSolution=FALSE$;

\textbf{Step2.} Choose $j$ uniformly at random among the nonnegative
integers not bigger than $m$.

\textbf{Step3.} Generate the initial state%
\begin{equation*}
|\Psi _{0}\rangle =\frac{1}{N}\sum\limits_{i=0}^{N-1}\sum%
\limits_{j=0}^{N-1}|\alpha \rangle |\beta \rangle \underbrace{|i\rangle
|j\rangle }|0\rangle |0\rangle |0\rangle |0\rangle
\end{equation*}

This can be achieved in $O(2\log _{2}N)$ steps using a $2\lceil \log
_{2}N\rceil -bit$ Hadamard transformation.

\textbf{Step4}. Apply $j$ iterations of $G_{DCT}^{\prime }$ starting from
the state $|\Psi _{0}\rangle $. i.e.,%
\begin{equation*}
|\Psi _{j}\rangle =(G_{DCT}^{\prime })^{j}|\Psi _{0}\rangle
\end{equation*}

\textbf{Step5.} Perform computation
\begin{equation*}
|\Psi _{end}\rangle =B_{inner}^{\prime }U_{L}^{\prime }|\Psi _{j}\rangle
\end{equation*}
\

\textbf{Step6.} Observe the 3rd, 4th and 7th register: let $(i_{0},j_{0})$
and coefficient $g_{i_{0}j_{0}}$ be the outcome respectively.

\textbf{Step7.} If $g_{i_{0}j_{0}}\in \{g_{ij}\mid \alpha \leq
(g_{ij})^{2}\leq \beta \}$, let $IsSolution=TRUE$, and \emph{exit};
otherwise, go to step 8.

\textbf{Step8.} Set $m$ to $\min \{\lambda m,N\}$ and go to step 2.

Similar to equation (28), Subroutine 2 has time complexity%
\begin{equation}
Time(Subroutine2)=O(\frac{N}{\sqrt{t}})t_{M}  \label{36}
\end{equation}

where $t$ is the numberf of elements in set $\{g_{ij}\mid \alpha \leq
(g_{ij})^{2}\leq \beta \}$.

\subsection{Quantum 2-D DCT (QDCT2)}

\begin{algorithm}
\textbf{Quantum Algorithm to compute matrix }$G$\textbf{\ approximately}
\end{algorithm}

\textbf{Input:} The line vectors of matrix of $D$, that are stored in
database with indices $i$ (Figure 10);

The column vectors of matrix of $F$, that are stored in database with
indices $j$ (Figure 10);

\textbf{Parameters: }$\Delta E$, $\alpha $, $\beta $, threshold value $%
\varepsilon >0$.

$nCount_{i}$ ($0\leq i<N$): Counting the repeated times of coefficient $%
c_{i} $ that is found when repeatedly applying subroutine 2.

$nMaxRepetition$: The maximum number of allowed repeated times of subroutine
2.

$nS$: The number of solutions that have been obtained.

\textbf{Output:} the big elements of $G$ and corresponding indices $%
(i_{0},j_{0},g_{i_{0}j_{0}})$, $(i_{1},j_{1},g_{i_{1}j_{1}})$, $\cdots $, $%
(i_{R},j_{R},g_{i_{R}j_{R}})$ such that$\frac{(g_{i_{0}j_{0}})^{2}+\cdots
+(g_{i_{R}j_{R}})^{2}}{||F||^{2}}\leq \varepsilon $. The other DCT
coefficients are set to zero (Table 1).

\textbf{Step0:} $\Delta E=||F||^{2}=\underset{i=0}{\overset{N-1}{\sum }}%
\underset{k=0}{\overset{N-1}{\sum }}(f_{ik})^{2}$; $\alpha =\frac{\Delta E}{%
N^{2}}$, $\beta =\Delta E$; $nS=0$; $nCount_{i}=0$, where $0\leq i<N$.

By equation (1) and (2), $||F||^{2}=||\overrightarrow{f_{0}}||^{2}+||%
\overrightarrow{f_{1}}||^{2}+\cdots +||\overrightarrow{f_{N-1}}||^{2}$. That
is, we can calculate $||\overrightarrow{f_{0}}||^{2}$, $||\overrightarrow{%
f_{1}}||^{2}$, ..., $||\overrightarrow{f_{N-1}}||^{2}$ one by one , then sum
up these values. Thus,
\begin{equation*}
Time(Step0)=Nt_{I}+(N-1)t_{A}+t_{D}\leq (5N+2)t_{M}+(N-1)t_{A}
\end{equation*}

while($\frac{\Delta E}{||F||^{2}}\geq \varepsilon $)

\{

\ \ \ \textbf{Step 1:} Apply subroutine 2 to find a coefficient $%
(i_{k},j_{k},g_{i_{k}j_{k}})$;

\ \ \ \textbf{Step 2:} If $IsSolution=TRUE$ and $nCount_{i}=0$,

\ \ \ \ \ \ \ \ \ then $\Delta E=\Delta E-(g_{i_{k}j_{k}})^{2}$, $nS=nS+1$; $%
\alpha =\frac{\Delta E}{N^{2}-nS}$, $\beta =\Delta E$; $%
nCount_{i_{k}}=nCount_{i_{k}}+1$;

\ \ \ \ \ \ \ \ \ \ If $IsSolution=TRUE$, and $%
nCount_{i_{k}j_{k}}<nMaxRepetition$, continue;

\ \ \ \ \ \ \ \ \ \ If $IsSolution=TRUE$, and $nCount_{i_{k}j_{k}}\geq
nMaxRepetition$, \textbf{exit} and apply classical algorithm to obtain DCT
coefficients.;

\ \ \ \ \ \ \ \ \ \ If $IsSolution=FALSE$, \textbf{exit} and apply classical
algorithm to obtain DCT coefficients.;

\}

\begin{algorithm}
\textbf{Quantum 2-D DCT (QDCT2)} Appling the above algorithm, we will get
the result of matrix $G$. Then compute 2-D DCT coefficients according to
equation $C=GD$ (i.e., equation (30)) by using the same method. That is,
applying the above method two times, we will obtain the 2-D DCT coefficients
of image.
\end{algorithm}

We can use classical method to compute the inverse DCT because many DCT
coefficients is equal to zero so that many running times will be saved.

Similar to equation (29), Quantum 2-D DCT (QDCT2) has time complexity:%
\begin{equation}
Time(QDCT2)=O(N)t_{M}  \label{37}
\end{equation}

In contrast, the classical 2-D DCT has time complexity $O(N^{2}\log _{2}N)$
multiplications. Even the efficiency classical parallel 2-D DCT has time
complexity $O(N\log _{2}N)+Time(Communication)$, where $Time(Communication)$
denotes the communication time between processors \cite{17}. Because many
data will be shared by all processors, the communication time will
drastically reduce the efficiency of parallel algorithm in fact \cite{18}.
By contrast, the there is no any communication time will be cost in Quantum
2-D DCT.

\section{The Comparison Between the Quantum DCT Iteration and the Grover
Iteration}

The quantum DCT iteration $G_{DCT}=(2|\xi \rangle \langle \xi
|-I)(U_{L})^{-1}(O_{inner})^{-1}O_{f}O_{inner}U_{L}$ is derived from Grover
iteration $G=(2|\Psi \rangle \langle \Psi |-I)O$. It includes two oracles;
by contrast Grover' has only one oracle. In classical computer, complex
computation must be decomposed into many simple function blocks (i.e.,
oracles) and then puts the function blocks back together to construct the
complex computation. This is a necessary processing because complexity
system is always constructed by many simple components The law is also
suitable to quantum computer maybe. Quantum DCT algorithm in this paper can
do two things, which are simultaneously calculating DCT coefficients and
simultaneously marking the wanted DCT coefficients. That is, complex
computation has been decomposed into two oracles $O_{inner}$ and $O_{f}$; by
contrast Grover's algorithm only can search a record with a given index.
Equation (25) is the key of applying two or more oracles to perform complex
computation. If there is no equation (25), applying two or more oracles is
no significant (Fig13)

The method presented in this paper generalizes Grover's algorithm. The
method is suitable to two or more arbitrary oracles. Clearly, if only if the
iteration includes the unitary with the form $(\cdots )^{-1}O_{f}(\cdots )$
such that state is eigenstate with eigenvalue $(-1)^{f(i)}$, the method is
valid.

The following table shows the difference between the quantum DCT iteration
and the Grover iteration.

\bigskip

\begin{tabular}{c}
{\Large Insert Table 2 Here}%
\end{tabular}

\bigskip

\section{Conclusion}

Discrete Cosine Transform (DCT) is very important in image compression.
Almost all of digital films such as VCD and digital pictures such as JPEG
files are compressed by utilizing DCT currently.

Real-time compressing and high compression ratio are the main research
topics of image compression. The classical 1-D DCT has complexity $O(N\log
_{2}N)$ for $N-dimensional$ vector. The time complexity of the classical 2-D
DCT is $O(N^{2}\log _{2}N)$ for $N\times N$ image. In general, the more $N$
is large, the more the compression ratio is high if the whole input data
sequence is Markov chain and the order of chain is bigger than $N$. However,
when $N$ is large, classical DCT does not satisfy the requirement of
real-time compressing. That is, finding a fast algorithm for large $N$ is
significant.

Quantum Fourier transform (QFT) and Klappenecker's Quantum DCT cannnot be
applied to image compression directly.

This paper presents the quantum DCT iteration $G_{DCT}$ that is defined as $%
G_{DCT}=(2|\xi \rangle \langle \xi
|-I)(U_{L})^{-1}(O_{inner})^{-1}O_{f}O_{inner}U_{L}$. And constructs 1-D DCT
based on the iteration with time complexity $O(\sqrt{N})$ for $N-dimensional$
vector and quantum 2-D DCT with time complexity $O(N)$ for $N\times N$
image. Two properties of DCT is utilized to design the quantum DCT algorithm
in this paper. One property is that DCT is energy conservation transform.
The other property is that many of the DCT coefficients have values close to
zero; these coefficients can be discarded without seriously affecting the
quality of the reconstructed image.

In classical computer, complex computation must be decomposed into many
simple function blocks (i.e., two or more oracles) and then puts the
function blocks back together to construct the complex computation. The law
is also suitable to quantum computer maybe. The quantum DCT iteration $%
G_{DCT}$ includes two oracles so that it can perform more complex search; by
contrast the Grover iteration $G=(2|\Psi \rangle \langle \Psi |-I)O$ has
only one oracle so that it can only perform simple search. Most quantum
algorithms have only one oracle currently, such as Shor's algorithm,
Grover's algorithm, Simon algorithm, and Deutsch-Jozsa algorithm.

$G_{DCT}$ is a rotation acting on subspace (Fig.14 and Fig15); by contrast,
the Grover iteration acts on global space. Acting on subspace reduce search
range drastically.

The method presented in this paper generalizes Grover's algorithm. Clearly,
if only if the iteration includes the unitary with the form $(\cdots
)^{-1}O_{f}(\cdots )$ such that state is eigenstate with eigenvalue $%
(-1)^{f(i)}$ (equation (25), Fig13), the method is valid. The method is
universal and can be applied to discrete Fourier transform and other
transforms for image compression.

\bigskip

\begin{acknowledgement}
We thank Professor Z.-F. Han for his encouragement to us. We thank Yong-Jian
Han, Pin-Xing Chen and Yun-Feng Huang especially for the significant
discussing with them. We would also like to thank Yong-Shen Zhang, Ke-Hui
Song, Guo-Ping Guo, Jian-Li, J.-M. Cai, M.-Y Ye, C. Han, Xiu -Ming Lin,
Yu-liang Li, Huai Ye, and other colleagues for their help. We thank Prof.
Jan-Zhang and Prof. Zhi-Lin Pu of Sichuan Normal Univ. for his help
\end{acknowledgement}

\section{List of Tables}

\textbf{Table 1:} The 2-D DCT coefficients of left-top $8\times 8$ blocks in
image (a): Discards all but 10 of the 64 DCT coefficients in each block, and
then reconstructs the image using the 2-D inverse DCT of each block.
Although there is some loss of quality in the reconstructed image, it is
clearly recognizable, even though almost 85\% of the DCT coefficients were
discarded. This property is utilized to design quantum algorithm in this
paper

\begin{center}
\begin{tabular}{|llllll|}
\hline
4.9211 & -0.00773 & 0.002581 & 0.001847 & $\cdots $ & 0 \\
0.014981 & -0.00281 & 0.002071 & 0 & $\cdots $ & 0 \\
0.008015 & 0.002014 & 0 & 0 & $\cdots $ & 0 \\
-0.0196 & 0 & 0 & 0 & $\cdots $ & 0 \\
$\vdots $ & $\vdots $ & $\vdots $ & $\vdots $ & $\vdots $ & $\vdots $ \\
0 & 0 & 0 & 0 & 0 & 0 \\ \hline
\end{tabular}
\end{center}

\textbf{Table 2}: The Comparison between quantum DCT iteration and the
Grover iteration

\begin{tabular}{|l|}
\hline
\begin{tabular}{c}
$G_{DCT}=(2|\xi \rangle \langle \xi
|-I)(O_{inner}U_{L})^{-1}O_{f}(O_{inner}U_{L})$; \\
${\small G=(2|\Psi \rangle \langle \Psi |-I)O}$%
\end{tabular}
\\ \hline
\begin{tabular}{c}
\begin{tabular}{l}
$(U_{L})^{-1}(O_{inner})^{-1}O_{f}O_{inner}U_{L}|\alpha \rangle |\beta
\rangle |i\rangle |\overset{\rightarrow }{f}\rangle |0\rangle |0\rangle
|0\rangle $ \\
$=(-1)^{f(i)}|\alpha \rangle |\beta \rangle |i\rangle |\overset{\rightarrow }%
{f}\rangle |0\rangle |0\rangle |0\rangle ${\small ;}%
\end{tabular}
\\
\multicolumn{1}{l}{${\small O|i\rangle =(-1)}^{f(i)}{\small |i\rangle }$}%
\end{tabular}
\\ \hline
\begin{tabular}{l}
$G_{DCT\text{ }}${\small has two oracles }$O_{inner}${\small \ and }$O_{f}$%
{\small ;} \\
$G${\small \ has only one oracle}%
\end{tabular}
\\ \hline
\begin{tabular}{l}
$G_{DCT\text{ }}${\small is a rotation on subspace;} \\
$G${\small \ is a rotation on global space.}%
\end{tabular}
\\ \hline
\begin{tabular}{l}
$G_{DCT}${\small \ can perform more complex search;} \\
$G${\small \ only can perform simple search}%
\end{tabular}
\\ \hline
{\small The method presented in this paper generalizes the Grover iteration.}
\\ \hline
\end{tabular}%
\bigskip

\section{List of Figures}

\begin{figure}[tbp]
\includegraphics[height=4cm,width=8cm]{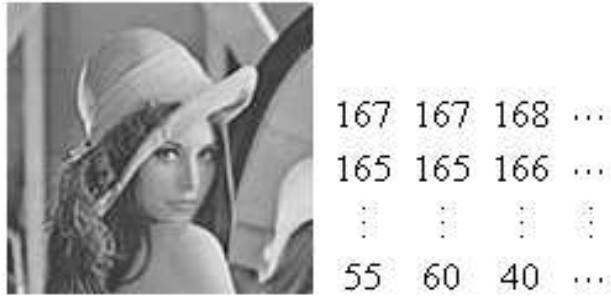}
\caption{The image with size $128\times 128$ is represented by the right $%
128\times 128$ matrix . Each element of this matrix is called \emph{pixel}.
The pixel value $f_{ij}$ is a gray-level value. The value $f_{ij}^{2}$ is
proportional to brightness or energy. The more $f_{ij}$ is large, the more
corresponding point in image is bright.}
\end{figure}

\begin{figure}[tbp]
\includegraphics[height=6cm,width=6cm]{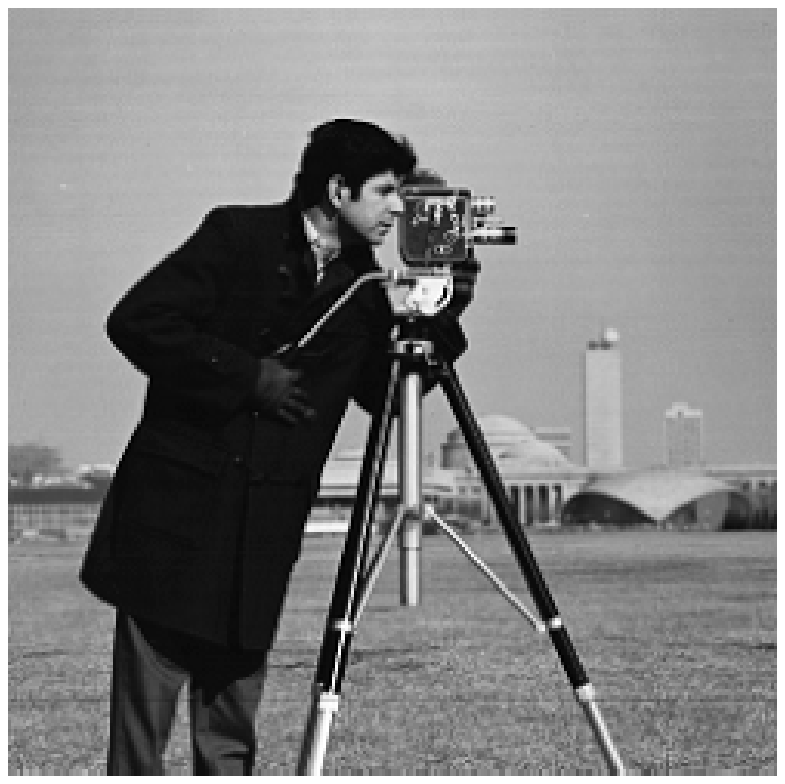}
\caption{$256\times 256$ image.}
\end{figure}

\begin{figure}[tbp]
\includegraphics[height=6cm,width=6cm]{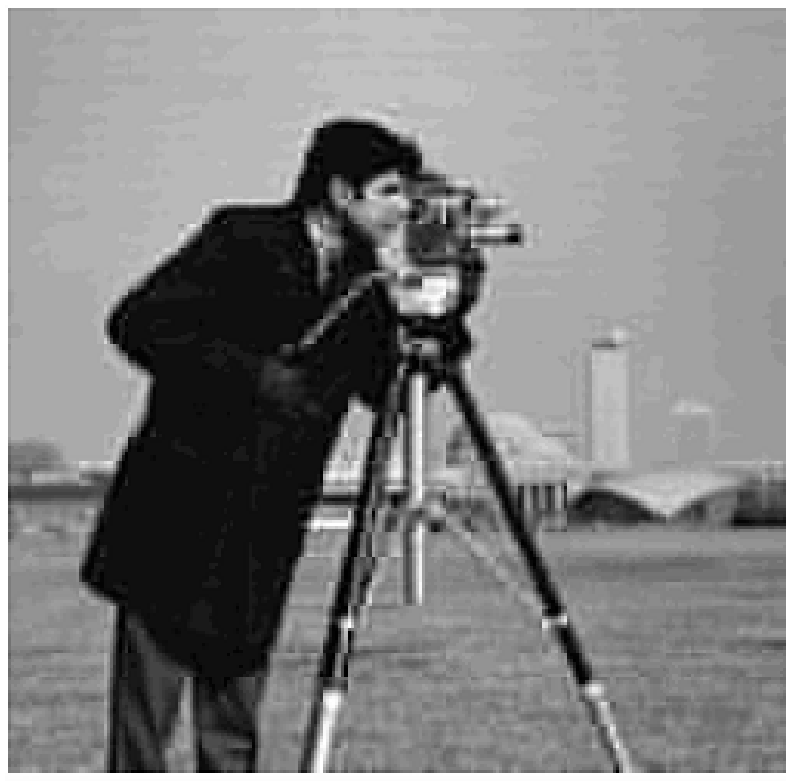}
\caption{reconstructed image}
\label{No02b}
\end{figure}

\begin{figure}[tbp]
\includegraphics[height=4.78cm,width=7.34cm]{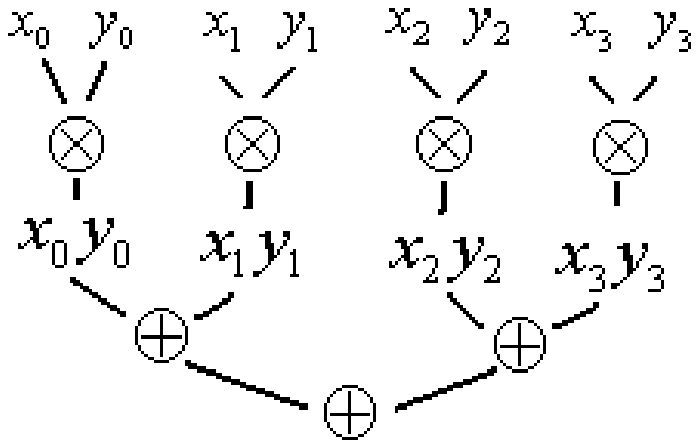}
\caption{The Parallel Circuit for Computing Inner Product: when $N\leq
2^{4\times 25}$, $t_{I}\leq 5t_{M}$}
\label{No3}
\end{figure}

\begin{figure}[tbp]
\includegraphics[height=5.31cm,width=7.13cm]{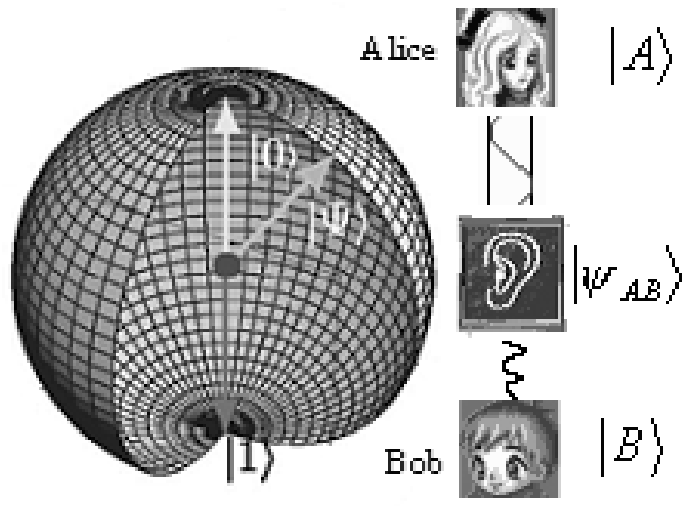}
\caption{The schematic diagram of the analogies between quantum
superpositions and sonic wave.}
\label{No4}
\end{figure}

\begin{figure}[tbp]
\includegraphics[height=3.63cm,width=7.13cm]{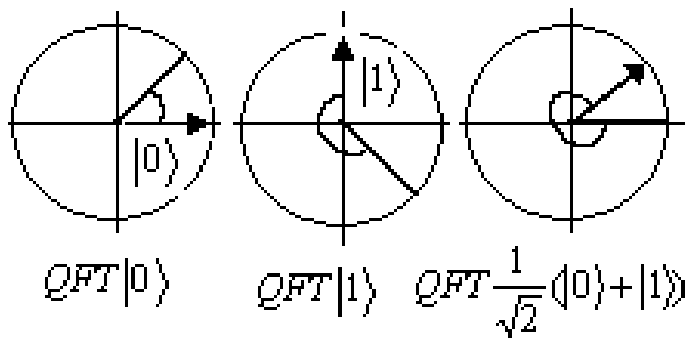}
\caption{$QFT$ only acts on basis states $|0\rangle $ and $|1\rangle $: $QFT%
\frac{1}{\protect\sqrt{2}}(|0\rangle +|1\rangle )=\frac{1}{\protect\sqrt{2}}%
(QFT|0\rangle +QFT|1\rangle )$}
\label{No5}
\end{figure}

\begin{figure}[tbp]
\includegraphics[height=3.88cm,width=4cm]{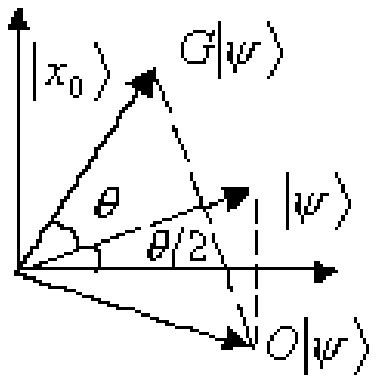}
\caption{The schematic diagram of rotating the state $|\Psi \rangle $ by $%
\protect\theta $ radians per application of $G$}
\label{No6}
\end{figure}

\begin{figure}[tbp]
\includegraphics[height=3.63cm,width=7.38cm]{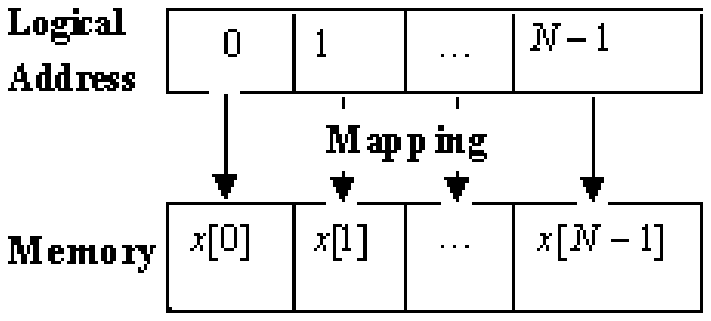}
\caption{The Conception of the Logical Mapping. The mapping associates data
with the corresponding logical address}
\label{No7a}
\end{figure}

\begin{figure}[tbp]
\includegraphics[height=5.31cm,width=9.25cm]{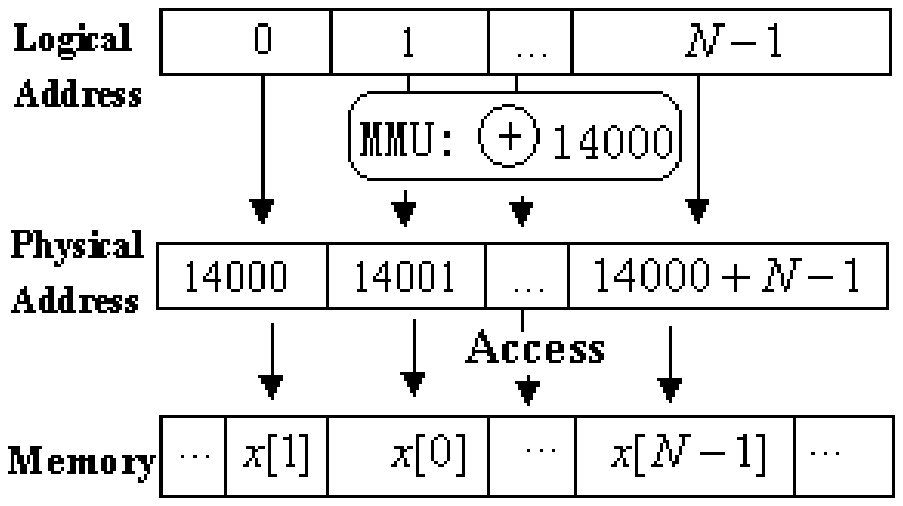}
\caption{The Illustration of the Physical Realization of the Logical Mapping
\protect\cite{16}. Accessing data is very very fast operation so that the
time of access can be ignored when designing algorithm. }
\label{No7b}
\end{figure}

\begin{figure}[tbp]
\includegraphics[height=5.38cm,width=8.47cm]{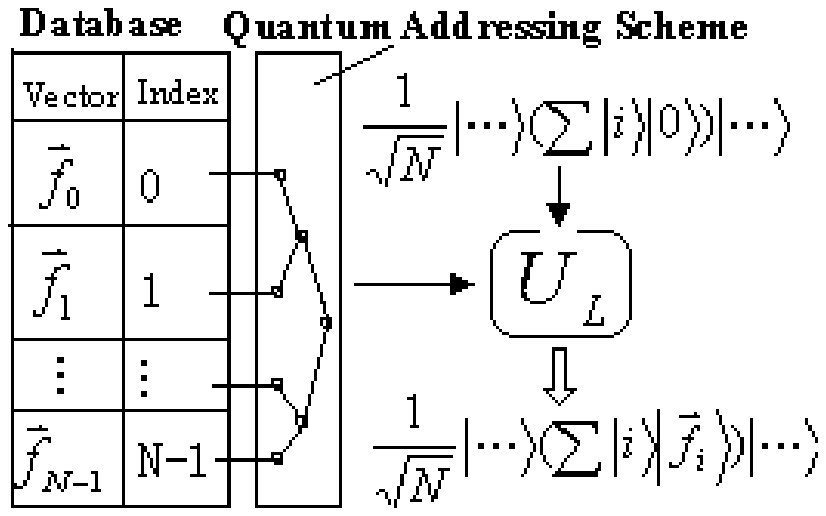}
\caption{The Representation of Image by Using Quantum States: LOAD operation
$U_{L}:$ $|i\rangle |0\rangle \protect\overset{U_{L}}{\mapsto }|i\rangle
|0\oplus \protect\overset{\rightarrow }{f_{i}}\rangle $ is a CNOT operation
and is a very very fast operation so that the time of addressing can be
ignored when designing quantum algorithm such as Grover's algorithm. It is
clear that the most efficient possible algorithm is in this model of
computation}
\label{No8}
\end{figure}

\begin{figure}[tbp]
\includegraphics[height=4.13cm,width=7.19cm]{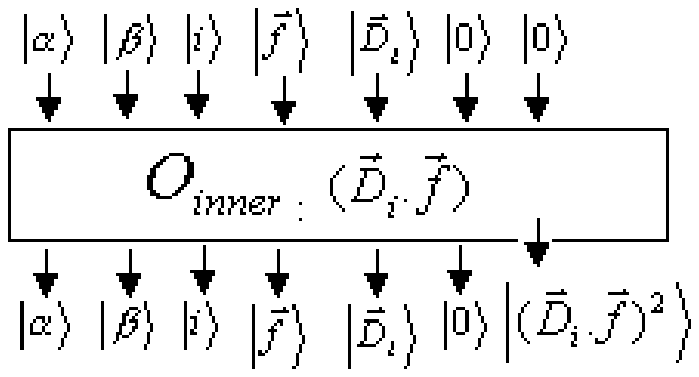}
\caption{The Conception of Oracle $O_{inner}$}
\label{No9}
\end{figure}

\begin{figure}[tbp]
\includegraphics[height=4.28cm,width=7.16cm]{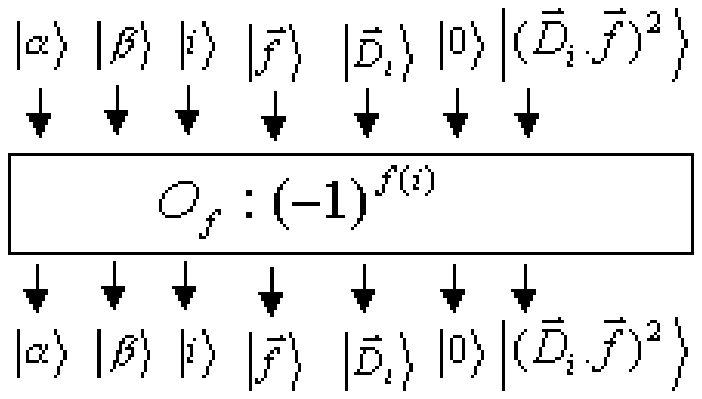}
\caption{The Conception of Oracle $O_{f}$}
\label{No10}
\end{figure}

\begin{figure}[tbp]
\includegraphics[height=4.13cm,width=7.19cm]{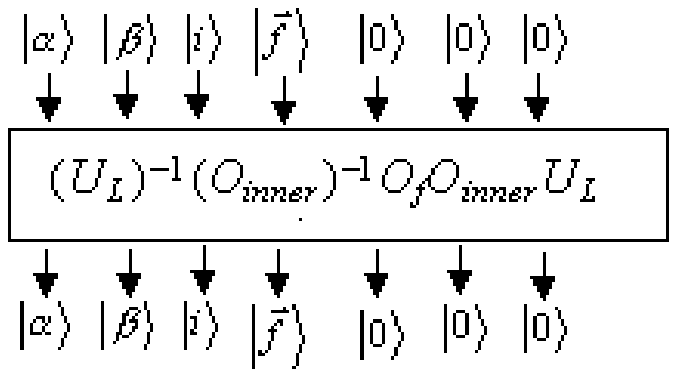}
\caption{$(U_{L})^{-1}(O_{inner})^{-1}O_{f}O_{inner}U_{L}$ only changes the
phase of the state as $(-1)^{f(i)}$. This is a key of applying two or more
oracles to perform complex search.}
\label{No11}
\end{figure}

\begin{figure}[tbp]
\includegraphics[height=8.25cm,width=9.63cm]{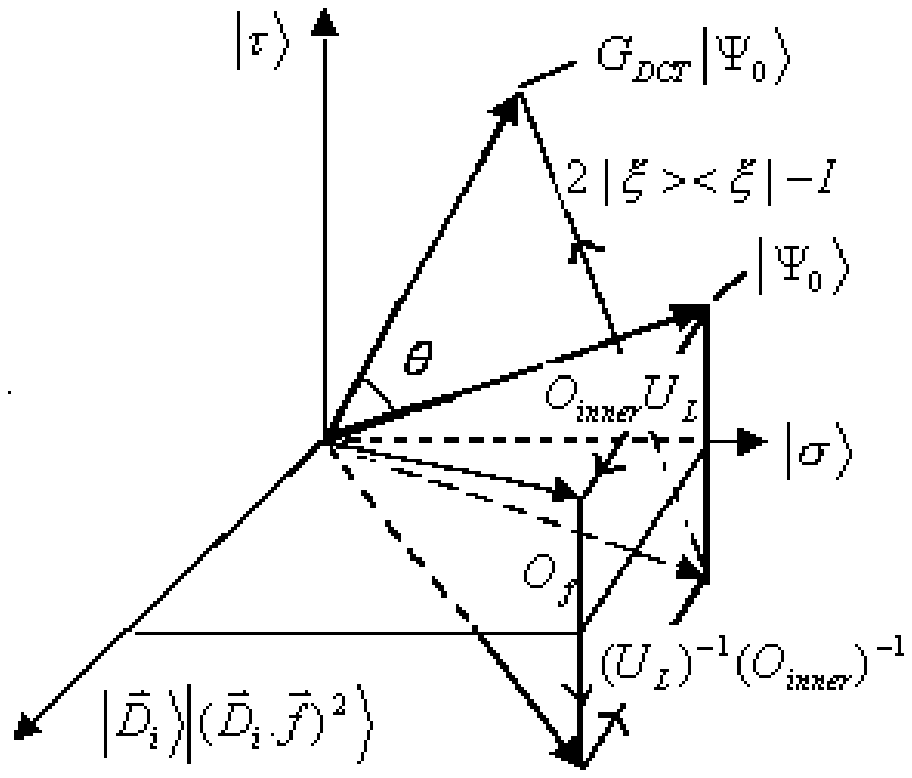}
\caption{The Action of A Single Quantum 1-D DCT Iteration $G_{DCT} $: The
initial state $|\Psi _{0}\rangle $ is rotated in the subspace $%
span\{|i\rangle \}$ by $\protect\theta $\ towards the superposition $|%
\protect\tau \rangle $\ of all solutions to the search. Initially, it is
inclined at angle $\frac{\protect\theta }{2}$\ from $|\protect\sigma \rangle
$, a state orthogonal to $|\protect\tau \rangle $. The product operation $%
(U_{L})^{-1}(O_{inner})^{-1}O_{f}O_{inner}U_{L}$ reflects the state about
the state $|\protect\sigma \rangle $, then the operation $2|\protect\xi %
\rangle \langle \protect\xi |-I$ reflect it about $|\protect\xi \rangle $\
in the subspace $span\{|i\rangle \}$.}
\label{No12a}
\end{figure}

\begin{figure}[tbp]
\includegraphics[height=4.41cm,width=7.97cm]{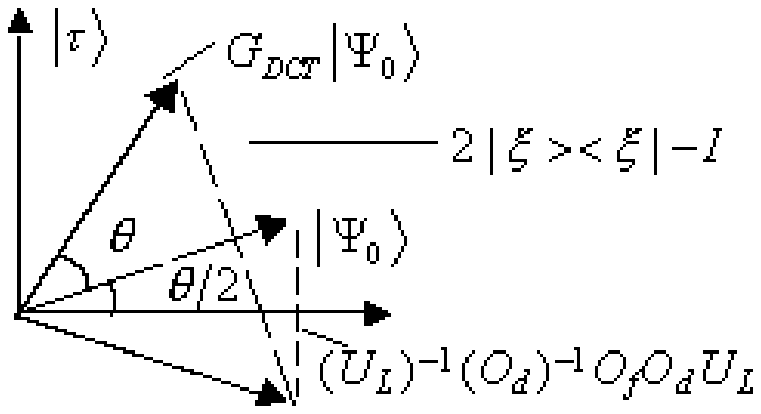}
\caption{$G_{DCT}$ is equivalent to a rotation on subspace $span\{|i\rangle
\}$}
\label{No12b}
\end{figure}

\begin{figure}[tbp]
\includegraphics[height=6cm,width=10cm]{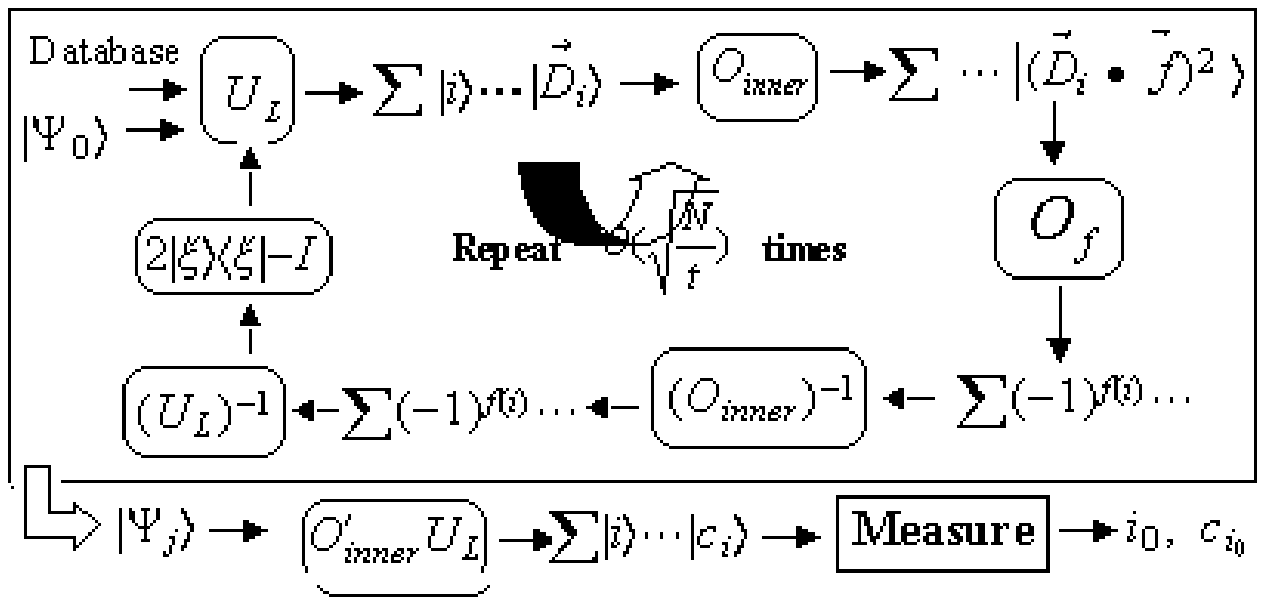}
\caption{Schematic processing of subroutine 1}
\label{No13}
\end{figure}

\begin{figure}[tbp]
\includegraphics[height=5cm,width=8cm]{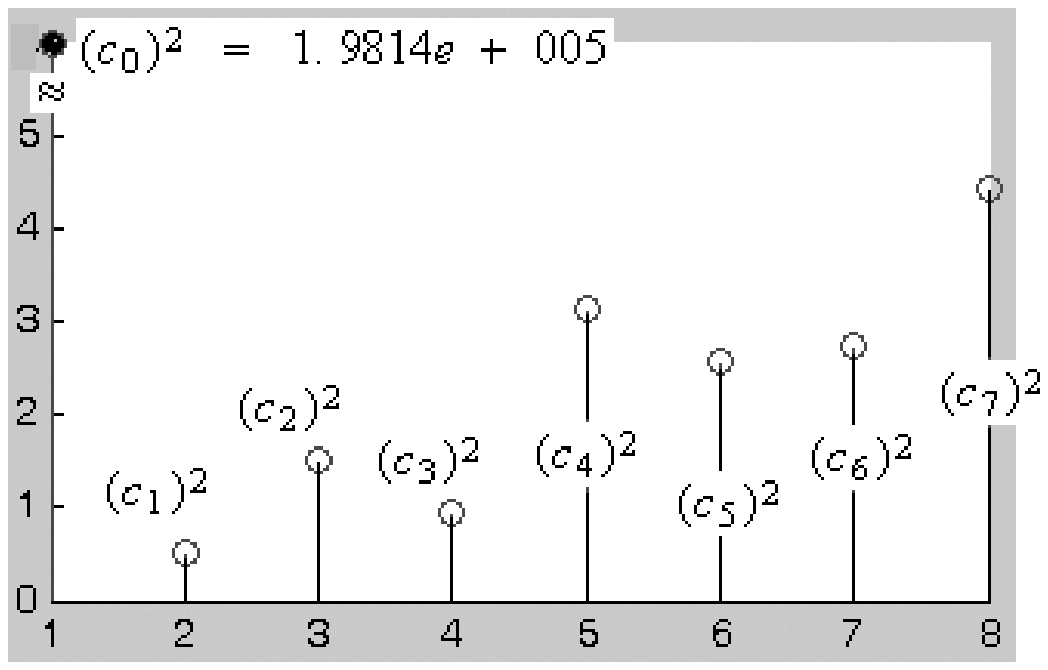}
\caption{Solution Set: $S=\{c_{0}\}$.The output is $(0,c_{0})$.}
\label{No14a}
\end{figure}

\begin{figure}[tbp]
\includegraphics[height=5cm,width=8cm]{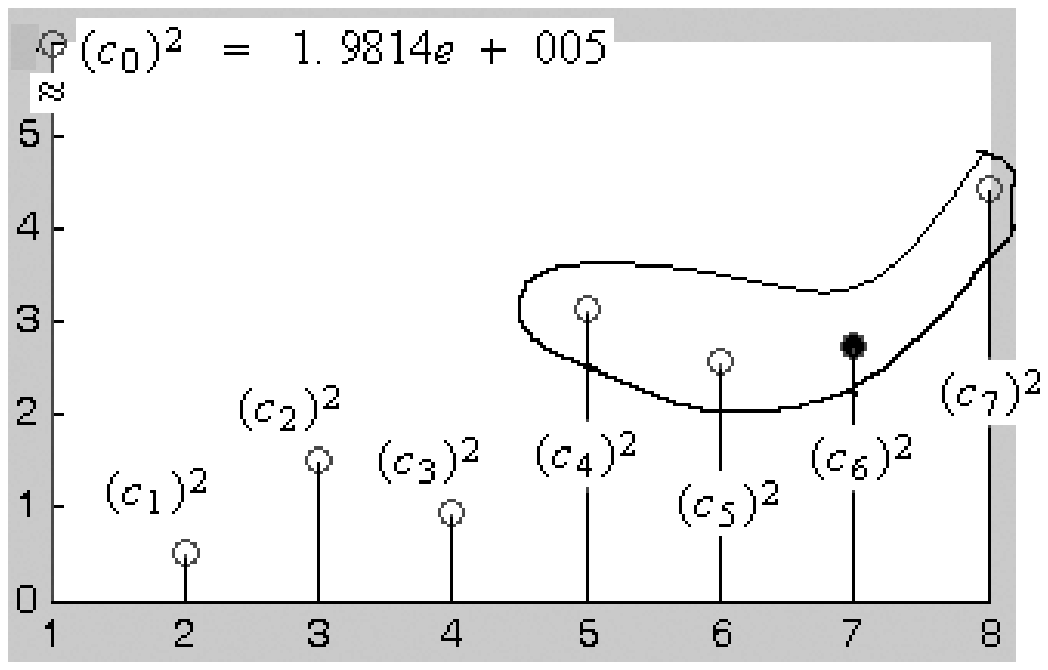}
\caption{Solution Set $S=\{(4,c_{4}),(5,c_{5}),(6,c_{6}),(7,c_{7})\}$.We
will obtain one of solution with equal probability.The output is $(6,c_{6})$}
\label{No14b}
\end{figure}

\end{document}